\documentclass[11pt,a4paper]{article}
\usepackage[T1]{fontenc}
\usepackage[utf8]{inputenc}
\usepackage{authblk}
\usepackage{graphicx} 
\usepackage[T1]{fontenc}
\usepackage{amsmath,amssymb}
\usepackage{amsthm}
\usepackage{authblk}
\usepackage{mathrsfs}
\usepackage{appendix}
\numberwithin{equation}{section}
\usepackage[bookmarksnumbered=true]{hyperref} 
\usepackage{amscd}
\usepackage{float}
\hypersetup{
    colorlinks = true,
    linkcolor = blue,
    anchorcolor = blue,
    citecolor = blue,
    filecolor = blue,
    urlcolor = blue
    }

\def\RR{\mathbb{R}}
\def\ZZ{\mathbb{Z}}

\def\eps{\varepsilon}

\def\Pell{\mathcal{P}} 

\title{Properties of the Ammann-Beenker tiling and its square periodic approximants}
\author[1]{Anuradha Jagannathan }
\author[2]{Michel Duneau }
 \affil[1]{Laboratoire de Physique des Solides, Universit\'e Paris-Saclay, 91405 Orsay, France}
 \affil[2]{ Centre de Physique Th\'eorique, CNRS, Ecole polytechnique, Institut Polytechnique de Paris, 91120 Palaiseau, France}

\date{December 2023}
\begin{document}
\maketitle

\begin{abstract}
Our understanding of physical properties of quasicrystals owes a great deal to studies of tight-binding models constructed on quasiperiodic tilings. Among the large number of possible quasiperiodic structures, two dimensional tilings are of particular importance -- in their own right, but also for information regarding properties of three dimensional systems. We provide here a users manual for those wishing to construct and study physical properties of the 8-fold Ammann-Beenker quasicrystal, a good starting point for investigations of two dimensional quasiperiodic systems. This tiling has a relatively straightforward construction. Thus, geometrical properties such as the type and number of local environments can be readily found by simple analytical computations. Transformations of sites under discrete scale changes -- called inflations and deflations -- are easier to establish compared to the celebrated Penrose tiling, for example.  We have aimed to describe the methodology with a minimum of technicalities but in sufficient detail so as to enable non-specialists to generate quasiperiodic tilings and periodic approximants, with or without disorder. The discussion of properties includes some relations not previously published, and examples with figures.
\end{abstract}
\section{Background and motivation}
Since the discovery of quasicrystals, there have been many questions as to their physical properties, and in particular, how these might differ from the properties of periodic crystals. In particular, concerning electronic properties, a fruitful approach consists of studying  models constructed on quasiperiodic tilings \cite{grunbaum}. Among such structures, works on one-dimensional (1D) quasicrystals are the most abundant, due to their comparative tractability -- both of theoretical analyses as well as of numerical computations in very large systems. Two dimensional tilings have also been quite extensively investigated and among these, the Penrose tiling and the Ammann-Beenker (aka octagonal) tiling shown in Fig.1, have been the two most frequently investigated. The reasons for the ubiquity of 2D tilings in theoretical investigations are multiple: firstly, one can study the effects of the quasiperiodicity in models which are parameter-free \footnote{that is, upto a single overall energy scale} and solely dependent on geometry. This is in contrast to one dimension, where it is necessary to introduce additional parameters to describe the quasiperiodic modulation -- as for example in the Fibonacci chain models, where two different hopping amplitudes -- or two onsite energies -- must be introduced. Secondly, new phenomena which are absent in 1D can arise, such as weak localization of electronic states. Quite large systems can be considered in numerical simulations, making it possible to extrapolate to the infinite 2D quasicrystal. Moreover, properties studied in 2D systems can often help predict what to expect for the physics of more complex (usually) 3D systems.  Modelling very large 3D quasiperiodic systems of course remains a goal, but it has not thus far been feasible to study sufficiently large system sizes, for reasons of computer time and memory. The Ammann-Beenker tiling which we will focus here on is particularly interesting, not only because there exist real quasicrystals with eight-fold symmetry \cite{wangchen}, but more importantly because, as we will discuss here, it is a good paradigm for quasiperiodically ordered systems.
 
A large variety of models have been introduced for studies of electronic properties on tilings. Typically, one considers tight-binding Hamiltonians in which electronic orbitals are localized on vertices of the tiling, and which are of the form 
\begin{equation}
H = \sum_{\langle i,j \rangle} \sum_{\sigma} t_{ij} (c^\dag_{i,\sigma} c_{j,\sigma} + h.c.)  + \sum_{i,\sigma} E_ic^\dag_{i,\sigma}c_{i,\sigma} + H_{int} 
\label{eq:ham1}
\end{equation}
where $i$ and $j$ are indices of sites (vertices), $\sigma$ is the spin index, and the $t_{ij}$ represent hopping amplitude strength (here taken to be independent of spin direction) along the bond between sites $i$ and $j$. $E_i$ are site-dependent energy parameters. $H_{int}$ represents the interacting part -- it could include  interactions between electrons or between electrons and impurities, etc. The above family of models include perfect deterministic tilings and also disordered ones. Disorder can be introduced rather simply in several ways: via the parameters of the Hamiltonian above or through random permutations of tiles. These model Hamiltonians are useful, firstly, to capture qualitatively properties of real laboratory systems. They are, secondly, also interesting in their own right, especially since it has now become possible to realize Hamiltonians such as Eq.\ref{eq:ham1} using cold atoms, and study them by experiments, as discussed in \cite{schneider2023}.

In perfect tilings, studies of non-interacting models
in one, two or three spatial dimensions have shown that quasiperiodicity tends to result in quite complex electronic properties as compared to those of periodic or amorphous systems (see the review by Grimm and Schreiber \cite{grimmreview}). These properties can be considered to arise from quantum interference effects in a quasicrystal, which can lead to a hierarchy of power laws both in space and time. They are  reflected in the appearance of singularities in the density of states as a function of energy. Such singular properties are particularly strong in 1D where they have been extensively investigated, especially in the Fibonacci chain which has served as a paradigm for quasicrystals (reviewed in \cite{jagaRMP}). Not surprisingly, for 2D tilings, there are far fewer exact results for the spectrum and wavefunctions. One notable result concerns the explicit construction of 
the ground states of the hopping model on the Ammann-Beenker and the Penrose tilings  \cite{kalugin2014}. These states have been shown to be multifractal (or critical) states, whose fractal dimensions can be computed using the geometrical structure of the tilings \cite{mace2017}. Geometrical considerations also enter in the classification and counting of degeneracies of confined states --  localized states with exactly zero amplitude everywhere except on a finite set of sites -- which exist for certain specific energies. The set of confined states which arise for $E=0$ in pure hopping models have been enumerated exactly in several 2D tilings \cite{koga,mirza,oktel}.

\begin{figure} [H] 
\centering
\includegraphics[height=200pt] {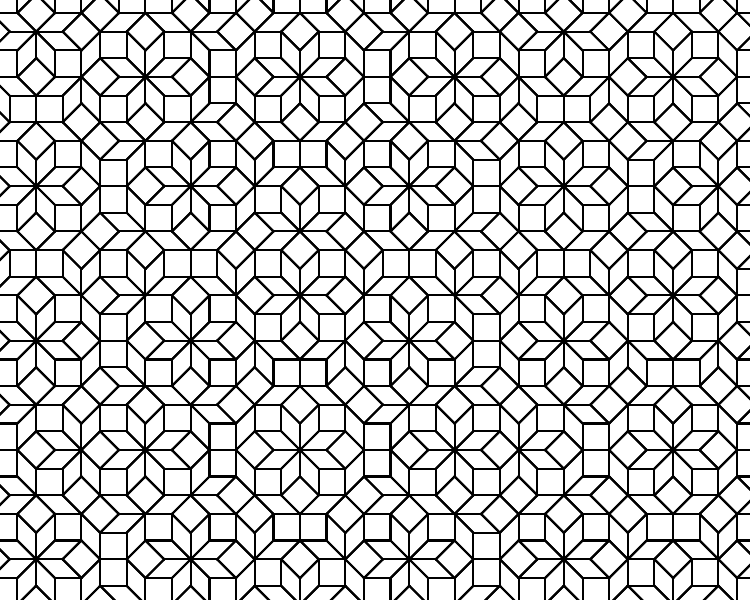} 
 \caption{A piece of the Ammann-Beenker tiling showing its construction from square and 45$^\circ$ rhombus tiles. Notice that in this tiling, at the level of first nearest neighbors, there are only six local environments (upto 8-fold rotations).}
\label{fig:mace}
\end{figure}

A large number of approximate results of electronic properties have been obtained by numerical calculations. These studies indicate that, in perfect tilings, the typical single electronic state is a power-law extended -- or critical -- state. These states are markedly different from the Bloch states of periodic crystals, and also from the exponentially decaying states of strongly disordered systems. On the one hand, the absence of extended states à la Bloch is understandably due to absence of translation invariance. On the other hand, the relatively slow (power-law) decay of wavefunctions compared say with those of random systems can be understood as well. This follows from the fact that quasicrystals possess a weak form of translation invariance in the following sense : given a patch of any size, one is guaranteed to find an identical patch within a small distance of the order of the patch radius. This property, called the Conway theorem, distinguishes them from random structures and makes unlikely Anderson-type localized states. Electron Green's functions in quasiperiodic tilings are therefore typically long-ranged, as are response and correlation functions. Addition of disorder, be it in the geometrical properties (in phason disordered tilings, as  discussed later) or in energy parameters of  Eq.1, can lead to new qualitatively different properties, with a distinction to be made between weak and strong disorder regimes \cite{grimmroemer,tarzia}. 

In many numerical studies, models are preferentially studied on approximant tilings, which have the property of repeating periodically, so that periodic boundary conditions can be applied. Arbitrarily large approximants can be constructed, and they tend towards the perfect quasicrystal in the limit of infinite size.  For the A-B tiling, a series of square approximants were introduced in \cite{duneau89,duneau94}. A  discussion of approximants of the octagonal tiling and the wider class of so-called Beenker tilings is given in the papers by Bedaride and Fernique \cite{bedaride}.

The advantages of working with approximants are multiple. Firstly, of course, boundary effects such as localized states at edges can be avoided. Approximants also allow one to estimate finite size corrections to quantities in a systematic fashion, and extrapolate to the quasiperiodic limit. In addition, the existence of an infinite series of periodic approximants is a crucial element of the gap labeling scheme for the integrated density of states \cite{aj2023}. This work showed that the gaps of the electronic spectrum of approximants can be labeled in terms of Pell numbers $Q_n$, defined in the Appendix A. In the infinite size limit, the special values of the integrated density of states tend towards the gap labeling theorem due to Kellendonk and Putnam \cite{kelput}. These labels, topological in origin, are robust under changes of the parameters of the model, as long as the corresponding gap remains open, and is a two-dimensional analogue of a well-known theorem for 1D systems due to Bellissard and coworkers \cite{bellissardgap}. 

Other electronic properties that have been studied in this tiling include interaction-induced magnetic or charge ordered phases
\cite{koga,hauck},  topological \cite{varjas,cao,ghadimi} and non-topological superconductivity \cite{takemori,andrade,fukushima}, spin Hall effect \cite{huangLiu}, Kondo screening of impurities \cite{deandrade}, hyperuniformity \cite{lin,sakai} and effects of disorder \cite{tarzia}, spreading of wavepackets \cite{trambly} etc, to mention only a few examples. 
Going beyond electronic properties, tilings are also of interest in many other contexts such as Heisenberg spin models \cite{wessel,szallas2007}, dimer covering models \cite{ghosh}, and the coloring problem \cite{flicker}, to cite just some examples. 

To conclude this introduction, this ``users manual" is intended to provide a quick guide for non-specialists seeking to study their favorite models on these structures, and for those who wish to simply know more about them. We assume that readers are familiar with basic notions of quasiperiodic tilings, and if not, we refer them to the excellent books by Senechal \cite{senechal}, Stauffer and Deloudi \cite{deloudi}, and at a more advanced mathematical level, the book by Baake and Grimm \cite{baakegrimm}. 
We will present the Cut-and-Project (CP) method to obtain geometrical properties of the Ammann-Beenker tiling and its square approximants. The quasiperiodic tiling is discussed in Sec.2, where we give some of the main results. Sec.3 discusses the square approximants introduced by Duneau et al \cite{duneau89}, by a straight-forward modification of the CP method used for the quasiperiodic case. The Appendix A describes some useful properties of the Pell numbers. Appendix B describes a second method of obtaining square approximants that preserves the tile shapes -- all squares and 45$^\circ$ rhombuses have equal side lengths, as in the quasicrystal.

\section{The Ammann-Beenker tiling.}
In the CP method, selected points of a 4D hypercubic lattice $\ZZ^{4}$ are projected onto the physical plane $E$ -- the plane in which the tiling resides. The orthogonal directions to this plane define the perpendicular plane $E'$. The four unit vectors of the hypercubic lattice, which form an orthonormal basis, are denoted by $\vec{\varepsilon}_\mu$ with $\mu=0,1,2,3$. 

To determine the parallel and perpendicular planes, one begins by finding a representation of the octagonal symmetry group in 4D which leaves invariant the hypercubic lattice. One such generator is given by
\begin{align}
\label{Mmatrix}
g= \begin{pmatrix}
0 & 0 & 0 & -1 \\
1 & 0 & 0 & 0  \\
0 & 1 & 0 & 0  \\
0 & 0 & 1 & 0 \\
\end{pmatrix} .
\end{align}

\begin{figure} [H] 
\centering
\includegraphics[height=80pt] {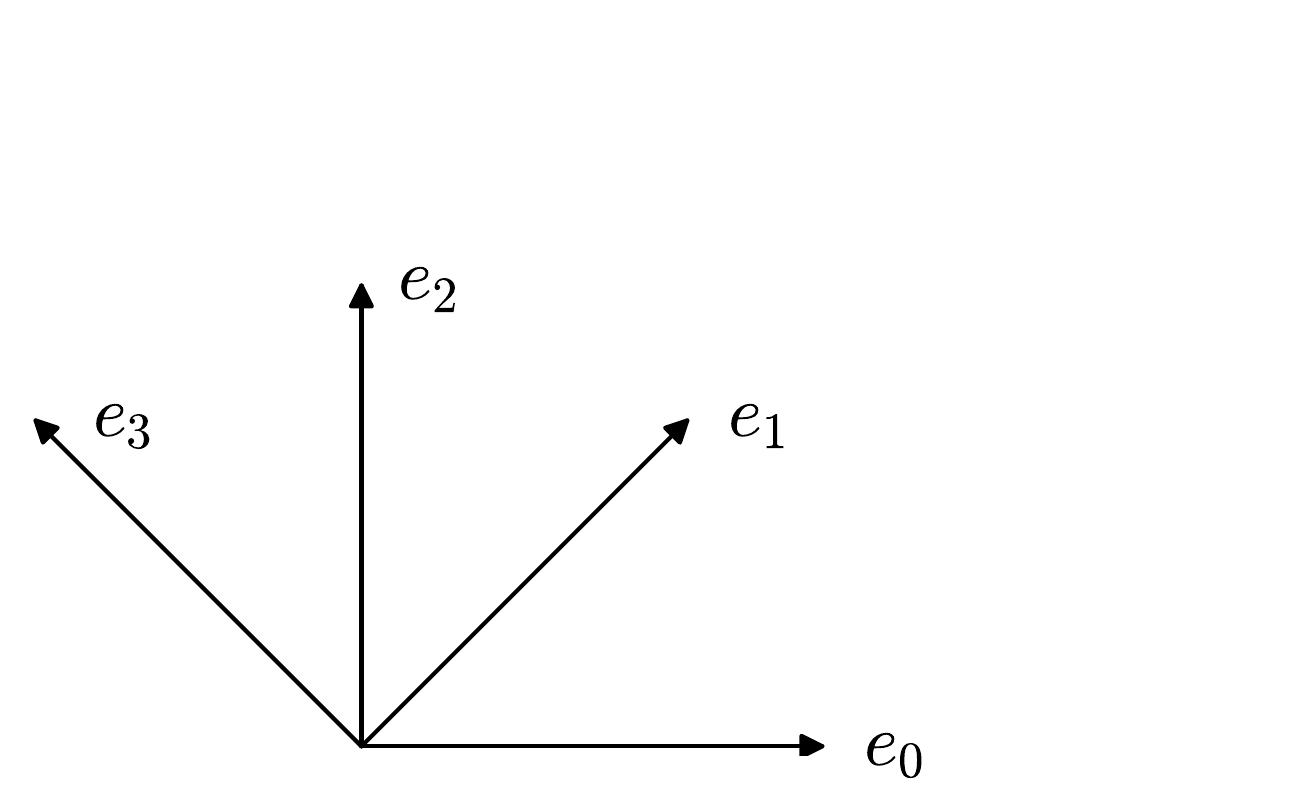} \hskip 1.0cm \includegraphics[height=120pt] {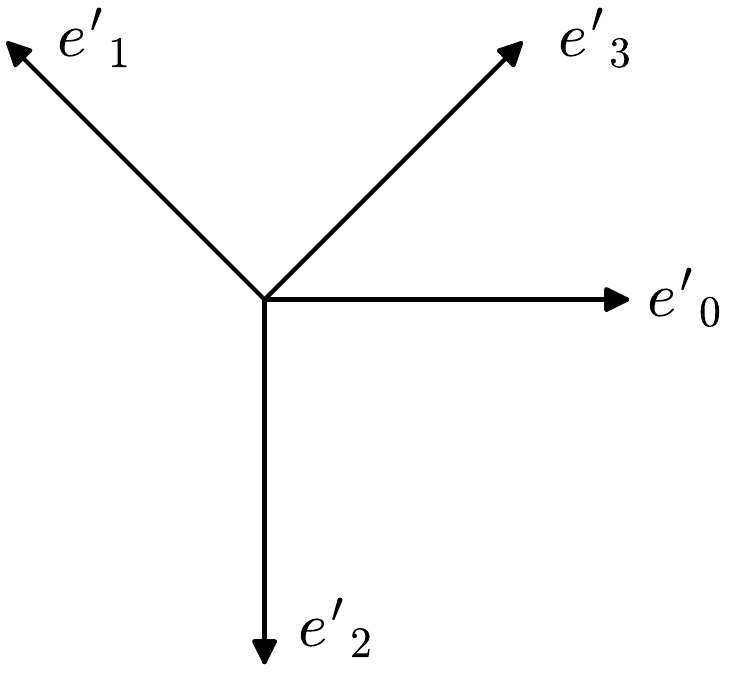}
\caption{The projections of the basis vectors $\vec{\eps}_\mu$ in $E$ (left) and $E'$ (right).}
\label{fig:basis}
\end{figure}

This representation can be reduced to two 2D representations in two orthogonal planes $E$ and $E'$, where $g$ is simply a 
rotation by $\pi/4$ and $3\pi/4$ respectively. The projection matrices $p$ and $p'$ from 4D onto these invariant planes $E$ and $E'$ are given by a classic formula involving characters of the representation which yields
\begin{align}
\label{projectors}
\begin{split}
p &:= \frac 1 4 \sum_{k=0}^7 \cos(k \tfrac \pi 4) g^k \ = 
\frac 1 2 \begin{pmatrix}
1 & \frac{\sqrt{2}} 2 & \phantom{-}0 & -\frac{\sqrt{2}} 2 \\
\frac{\sqrt{2}} 2 & 1 & \frac{\sqrt{2}} 2 & 0  \\
0 & \frac{\sqrt{2}} 2 & 1 & \frac{\sqrt{2}} 2  \\
-\frac{\sqrt{2}} 2 & 0 & \frac{\sqrt{2}} 2 & 1 \\
\end{pmatrix} , \\
p' &:= \frac 1 4 \sum_{k=0}^7 \cos(k \tfrac {3\pi} 4) g^k = 
\frac 1 2 \begin{pmatrix}
1 & -\frac{\sqrt{2}} 2 & 0 &  \frac{\sqrt{2}} 2 \\
-\frac{\sqrt{2}} 2 & 1 & -\frac{\sqrt{2}} 2 & 0  \\
0 & -\frac{\sqrt{2}} 2 & 1 & -\frac{\sqrt{2}} 2  \\
\frac{\sqrt{2}} 2 & 0 & -\frac{\sqrt{2}} 2 & 1 \\
\end{pmatrix} .
\end{split}
\end{align}
These projections matrices satisfy $p^2=p$, $p'^2=p'$, operating in $\RR^4$, and the projections of the 4D basis vectors onto $E$ and $E'$ are  $\vec{e}_i = p\vec{\eps}_i$, $\vec{e}'_i = p'\vec{\eps}_i$ for $0\leq i \leq 3$ with 
$\|\vec{e}_i\| = \|\vec{e}'_i\| = \tfrac1{\sqrt{2}}$ as shown in Fig.\ref{fig:basis}. Once 2D bases are chosen, such as 
$\{e_0,e_2\}$ in $E$ and $\{e'_0,-e'_2\}$ in $E'$, it is convenient to represent $p$ and $p'$ by the $2\times 4$ matrices $\pi$ and $\pi'$ given in \ref{projectors2d}.\\ 

The $x,y$ (resp. $x',y'$) axes in parallel and perpendicular space are chosen as follows: in $E$, they are directed along $\{e_0,e_2\}$ corresponding to columns $1$ and $3$ of $p$. In $E'$ the axes are given by $\{e'_0,-e'_2\}$ corresponding to columns $1$ and $3$ of $p'$ : 
\begin{align*}
E  : e_0 &= \frac 12 (1 ,\phantom{-}\tfrac{\sqrt{2}}2 , 0 , -\tfrac{\sqrt{2}}2) , 
\quad e_2 = \frac 12 (0 ,\phantom{-} \tfrac{\sqrt{2}}2 , 1 ,\phantom{-} \tfrac{\sqrt{2}}2) ,\\
E' : e'_0 &= \frac 12 (1 ,-\tfrac{\sqrt{2}}2 , 0 , \phantom{-} \tfrac{\sqrt{2}}2 ) , 
\quad e'_2 = \frac 12 (0 , -\tfrac{\sqrt{2}}2 ,1 , -\tfrac{\sqrt{2}}2) .
\end{align*}
We turn next to the projection matrices $\pi$ and $\pi'$ which give the projections of a point of the hypercubic lattice, $\vec{R}$ onto $E$ and $E'$. 
To determine these, we note that the set $\{\vec{e}_0,\vec{e}_2,\vec{e}'_0,\vec{e}'_2\}$ is an orthogonal basis for $\RR^4$. Let $B$ be the 4 x 4 matrix 
$\{e_0,e_2,e'_0,e'_2\}$ and let $B^{-1}$ be its inverse. We have
\begin{align}
\label{Bmatrix}
B= 
\frac12
\begin{pmatrix}
1 & 0 & 1 & 0 \\
\frac{\sqrt{2}}2 & \frac{\sqrt{2}}2 & -\frac{\sqrt{2}}2 & -\frac{\sqrt{2}}2  \\
0 & 1 & 0 & 1 \\
-\frac{\sqrt{2}}2 & \frac{\sqrt{2}}2 & \frac{\sqrt{2}}2 & -\frac{\sqrt{2}}2 \\
\end{pmatrix} ,
B^{-1}= 
\begin{pmatrix}
1 & \frac{\sqrt{2}}2 & 0 & -\frac{\sqrt{2}}2 \\
0 & \frac{\sqrt{2}}2 & 0 & \frac{\sqrt{2}}2  \\
1 & -\frac{\sqrt{2}}2 & 0 & \frac{\sqrt{2}}2 \\
0 & -\frac{\sqrt{2}}2 & 1 & -\frac{\sqrt{2}}2 \\
\end{pmatrix} .
\end{align}

It is now simple to obtain the 2 x 4 matrices that map a point of the 4D hypercubic lattice onto $E$ and $E'$. Positions in the physical plane $E$ are given by $\vec{r}=\pi \vec{R}$ and those in the perpendicular plane $E'$ by  $\vec{r}'=\pi'\vec{R}$, where the matrices  $\pi$ and $\pi'$  are given by the first and last two rows respectively of $B^{-1}$. That is, 
\begin{align}
\label{projectors2d}
\pi &=  \begin{pmatrix}
1 & \frac1{\sqrt{2}} & 0 & \frac{-1}{\sqrt{2}}  \\
 0 & \frac1{\sqrt{2}} & 1 & \frac1{\sqrt{2}} 
\end{pmatrix} ,
&\pi' &= \begin{pmatrix}
1 & \frac{-1}{\sqrt{2}} & 0 & \frac1{\sqrt{2}} \\
0 & \frac{1}{\sqrt{2}} & -1 & \frac{1}{\sqrt{2}}  \\
\end{pmatrix} .
\end{align}
in the basis $\{e_0,e_2\}$ and $\{e'_0,-e'_2\}$ (where the minus sign is added to make the basis direct).

\subsection{Selection window}
A given point of $\ZZ^{4}$ is selected if it lies inside an infinite strip $\mathcal{S}$ that runs parallel to $E$. 
The selection window $W_0$ is the ``cross-section" of the strip, corresponding to the outer envelope of the projection of one unit cell of $\ZZ^{4}$ onto the perpendicular plane $E'$. One now defines a shifted window, $W=T({\vec{\delta}}) W_0$, where $T$ is a translation in perpendicular space. $W$ is simply the window $W_0$ translated by a shift vector $\vec{\delta}$. The CP method can be stated as follows: given a 4D lattice point $\vec{R}$, its perpendicular space projection
must lie within $W$, to be selected. If the condition is satisfied, then the point is projected in $E$, where it corresponds to a vertex of the tiling at the position $\vec{r} = \pi \vec{R}$. Inversely, one can lift a vertex of the tiling to 4D. The shift vector is arbitrary, it is advisable to take some generic choice of the shift vector $\vec{\delta}$, except when samples with specific symmetries, such as 8-fold rotational symmetry are desired (see Sec.2.4 below).

\begin{figure} [H] 
\centering
\includegraphics[height=64pt] {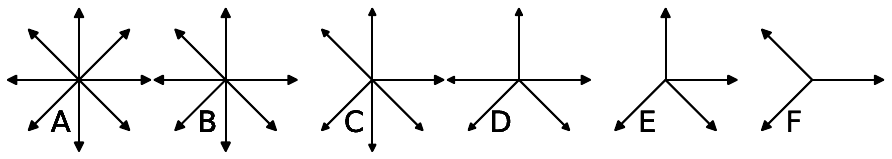}\\
\includegraphics[height=150pt] {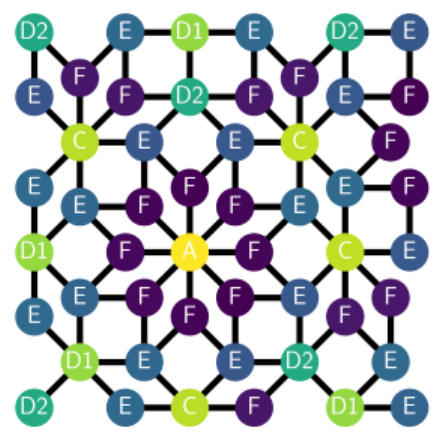}\\
\caption{Top: The vertex types (upto rotations) of the Ammann-Beenker tiling, labeled A (coordination number 8) to F (coordination number 3). Bottom: A patch of tiling with site types labeled. Note that there are two sub-categories of D sites (see text). Figure reproduced with permission from \cite{mace2017}.}
\label{fig:vertices}
\end{figure} 

\begin{figure} [H] 
\centering
\includegraphics[height=250pt] {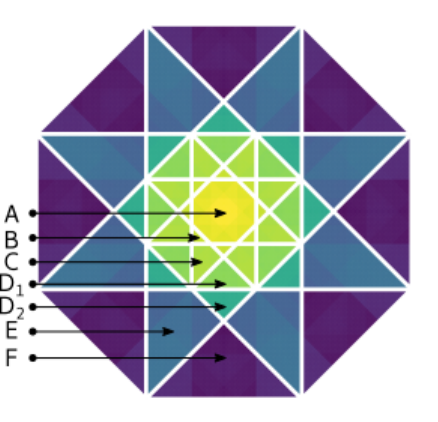}
\caption{The acceptance domains for the vertex types of Fig.\ref{fig:vertices}, labeled A (coordination number 8) to F (coordination number 3). Due to 8-fold symmetry only one domain of each type is labeled. Domains corresponding to two sub-categories of D sites have been shown (see text). Figure reproduced with permission from \cite{mace2017}. }
\label{fig:vertexperp}
\end{figure} 

\subsection{Vertex types and their frequencies}
Vertex types will be defined here by the nearest neighbor environment of each site. There are 6 distinct coordination numbers ($z$, the number of nearest neighbors) possible on the tiling. They range from $z=$8 (maximum) to $z=$3(minimum), and are labeled A,B,..,F  as shown in Fig.\ref{fig:vertices}. The reason for distinguishing two subcategories of the 5-fold D sites will be discussed in the next sub-section. It was shown in \cite{duneau89} that two vertices which could lie very far apart in real space, but which have the same nearest neighbor environment, must be projected onto the same region in $E'$. The window $W$ can thus be subdivided into six domains, labeled $A,..,F$ as shown in Fig.\ref{fig:vertexperp}. 

The central octagon $A$, colored yellow in Fig.\ref{fig:vertexperp}, is the region corresponding to sites of coordination number 8. The eight triangular regions marked $B$ correspond to vertices of coordination number 7. The  8 triangle-shaped domains $F$ correspond to sites of coordination 3, and so on. The 8-fold symmetry of the tiling holds in the perpendicular space representation as well, where the symmetry holds for each of the sub-domains individually. 

Since the entire window is uniformly and densely covered by projection of points of $\ZZ^4$  due to the irrational slope of $E'$, one can obtain the relative densities of different types of local environments by comparing the size of their domains in perpendicular space. A calculation of the areas subtended by each of the six families of sites yields the frequencies of occurrence of each vertex to be
\begin{align}
\{ f_A, f_B, ...., f_F \}  = \left\{ \frac{1}{\lambda}, \frac{2}{\lambda^2} , \frac{1}{\lambda^3} , \frac{2}{\lambda^4} , \frac{1}{\lambda^5} , \frac{1}{\lambda^4}\right\} 
\end{align}
where $\lambda = 1+\sqrt{2}$. 

The above argument can be extended to find the frequencies of local environments defined out to second nearest neighbors, third nearest neighbors, and so on. Each of these domains form sub-domains of smaller and smaller area. The mapping from real space clusters to perpendicular space domains was used by Oktel \cite{oktel} to compute the frequencies of different types of confined states in the AB tiling for the pure hopping model. 

\subsection{Inflation and deflation of the tiling}
\begin{figure} [H] 
\centering
\includegraphics[height=180pt] {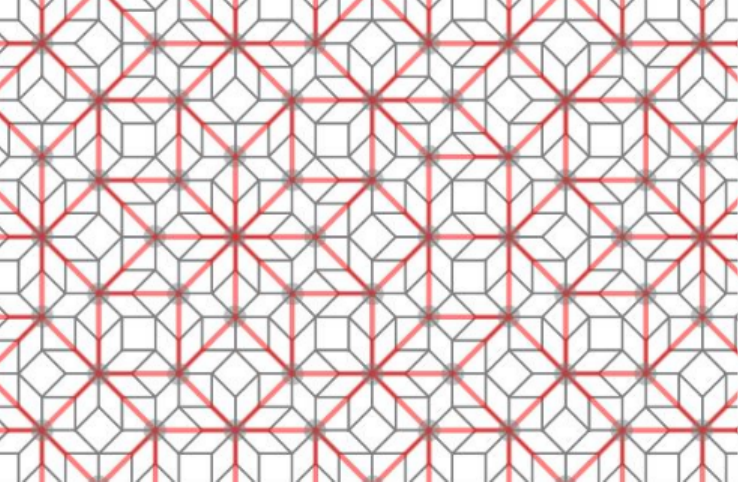}
\caption{A portion of the quasiperiodic tiling : the original tiling is shown by grey bonds. The inflated tiling is shown by red bonds which are longer than the grey bonds by a factor $\lambda=1+\sqrt{2}$. Large dots indicate sites that remain after inflation as vertices of the larger tiling.}
\label{fig:inflate1}
\end{figure}

\begin{figure} [H] 
\centering
\includegraphics[height=150pt] {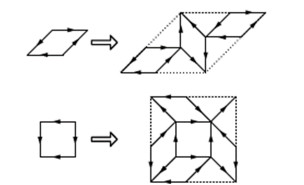} 
\caption{Inflation rules for the Ammann-Beenker tiling, showing the transformation of the rhombus and the square. }
\label{fig:inflationrules}
\end{figure}

Inflations and deflations are important symmetry transformations 
besides the octagonal rotations generated by \ref{Mmatrix}. They produce similar infinite octagonal tilings at different scales 
which contain the initial tiling (deflations) or are contained in it (inflations) as illustrated in Fig.\ref{fig:inflate1}. An alternative way to perform inflation or deflation operations is given by ``substitution rules" which define the passage from smaller tiles to bigger tiles (or vice-versa), as illustrated in Fig.\ref{fig:inflationrules}. The property of invariance under inflation has been exploited in theoretical treatments of electronic properties using the renormalization group \cite{sirebell,benza,zhongmoss}. 
The linear transformations invariant with respect to the octagonal group are of the form $x p + x' p'$, where $x,x'$ are numbers (real or complex) so that $E$ and $E'$ are invariant w.r.t. such transformations. Furthermore they must preserve the lattice $\ZZ^4$ so that they must be represented by unimodular $4\times 4$ matrices, which involves conditions bearing on $x$ and $x'$.

The inflation matrix $M$ is defined by
\begin{align}
\label{Mmatrix}
M = \lambda p+ \lambda'p' = \begin{pmatrix}
1 & 1 & 0 & -1 \\
1 & 1 & 1 & 0  \\
0 & 1 & 1 & 1  \\
-1 & 0 & 1 & 1 \\
\end{pmatrix} .
\end{align}
$M$ is an unimodular symmetric matrix which satisfies the relation $M^2 = 2M+ 1$, with eigenvalues are $\lambda$ 
and $\lambda'=1-\sqrt{2} = -\lambda^{-1}$, both doubly degenerate. The hypercubic lattice $\ZZ^{4}$ is invariant and the direct 
effect of $M$ is to scale up the set of vertices in $E$ by the factor $\lambda$, while scaling down the selection 
window in $E'$ by the factor $\lambda'$. \\
The $n$th power of $M$ satisfies the relation
\begin{align}
\label{eq:inflationmatrix}
M^n = \lambda^n p + \lambda'^n p'
\end{align}
where  $\lambda^n$ satisfies the recursion relation
\begin{eqnarray}
\lambda^n &=& \lambda\Pell_n  + \Pell_{n-1} \hskip 1cm ( n>0)  \nonumber \\
 &=& (-1)^{1-n} (\lambda\Pell_{-n}  - \Pell_{1-n}) \hskip 1cm( n<0) 
\end{eqnarray}
in terms of the Pell numbers, whose properties are briefly reviewed in the Appendix.

As can be seen from Eq.\ref{Mmatrix}, $M$ acts in real space to transform the original tiling of edge length $l_0=\frac{1}{\sqrt{2}}$ into one whose tile edge lengths are larger, $l_1=\lambda l_0$. In perpendicular space $M$ acts to decrease lengths by the factor $\lambda$. As a result the octagonal selection window of the new tiling is reduced by a factor $\lambda$ compared to $W$. 

Fig.\ref{fig:inflate1} shows a region of the tiling with bonds of the original tiling drawn in grey and those of the new tiling after inflation in red.  The sites which remain after inflation, becoming vertices of the new inflated tiling. These sites have perpendicular space coordinates which lie within an octagonal domain $W'$ which is smaller than $W$ by a factor $\lambda$. $W'$ is formed by the union of subdomains A,B,C and half of $D$, which is called $D_1$ (Fig.\ref{fig:vertexperp}). It can be observed that the D sites are of two types: those whose projections in $E'$ lie in $D_1$, and the others whose projections lie in $D_2$. In real space, the two types are distinguishable as well. This can be seen by inspecting their next nearest neighbor shells. Four of the nearest neighbors of $D_2$ sites are 3-fold coordinated F sites, whereas the nearest neighbors of $D_1$ are E-sites (two) and F-sites (two). The two kinds of D-site are found to occur as nearest neighbor pairs in the Ammann-Beenker tiling.

The classification of vertex types for sites which remain in the new inflated tiling --  are not decimated -- is easily carried out by inspecting the perpendicular space coordinates of the sites. To determine the new type, it suffices to divide the new window $W'$ into six subdomains, as was done for the original window. To give an example, it can be seen on inspecting Fig.\ref{fig:vertexperp} that if the site was of C-type before inflation then it must be an E-type after inflation. This follows from noting that the domain for E is exactly the domain for C after multiplication by a scale factor $\lambda$. Similarly, all $B$ sites are transformed under inflation to $D_2$ sites. Table 1 summarizes the initial and final site types before and after a single inflation. An A-site will remain A-type if its projection lies within an octagonal domain of edge length $1/\lambda$ times smaller than the original A-site domain (the yellow domain of Fig.4). These rules for transformation of sites were used in an renormalization group treatment of a Heisenberg model for spins on the Ammann-Beenker tiling \cite{jagannathanHAFM}. 

When the tiling is inflated twice, the sites which are left are those with projections within the central yellow octagon of edge $\lambda^{-2}$ in Fig.\ref{fig:vertexperp}. That is, the vertices of the twice-inflated tiling are just the ensemble of A-sites of the original tiling.
\begin{center}
\begin{tabular}{|l c r|}
\hline
Initial && Final  \\
\hline
A &\quad \quad $\rightarrow$ \quad \quad &A or B or C or $D_1$ \\
B &\quad \quad $\rightarrow$ \quad \quad & $D_2$ \\
C &\quad \quad $\rightarrow$ \quad \quad & E  \\
$D_1$&\quad \quad $\rightarrow$ \quad \quad & F \\
\hline
\end{tabular}
\vskip 0.5cm
\small{Table 1. Transformation of sites under inflation. In the first row of the table, the final site can be one of four types, depending on the perpendicular space coordinate of the initial A-site. }
\end{center}

An alternative method of generating the Ammann-Beenker tiling uses inflation rules. These are rules for transforming smaller tiles into larger ones, as shown in Fig.\ref{fig:inflationrules}. Notice that tile edges are decorated by arrows which have to be exactly matched at the boundary between two neighboring tiles. For more details see \cite{socolar}. One can compare these rules with the inflated tiling obtained by the CP methods shown in Fig.\ref{fig:inflate1}. 

\subsection{8-fold symmetric patches of the AB tiling}
\begin{figure} [H] 
\centering
\includegraphics[height=300pt] {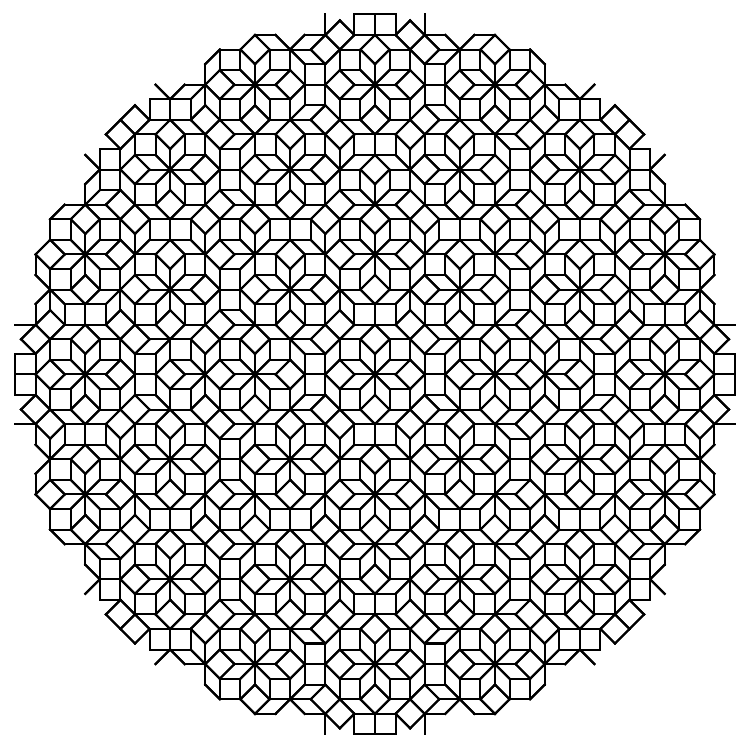} 
\caption{A finite sample (1185 sites) of the Ammann-Beenker tiling having a center of 8-fold rotational symmetry. }
\label{fig:symmetry8}
\end{figure}

For generic choices of the shift vector $\vec{\delta}$, patches generated by the CP method do not have a global center of rotation symmetry. However by choosing the shift arbitrarily close to the origin of $E'$, one can obtain bigger and bigger domains -- with, as the limiting case, an infinite tiling with a single point of perfect 8-fold symmetry at the origin. A finite patch (obtained )for a shift vector $\vec{\delta}=(0.003,0.001)$), and having a center of symmetry under 8-fold rotation at the origin is shown in Fig.\ref{fig:symmetry8}. Although not of interest for most studies of electronic properties \footnote{to the contrary, the presence of rotation symmetry introduces a huge redundance due to equivalence of sites under rotations.} these rotationally symmetric samples are used in studies of boundary-related phenomena, as in the topological phase studied by Varjas et al \cite{varjas} for example.

\subsection{Phasons}
\begin{figure} [H] 
\centering
\includegraphics[height=80pt] {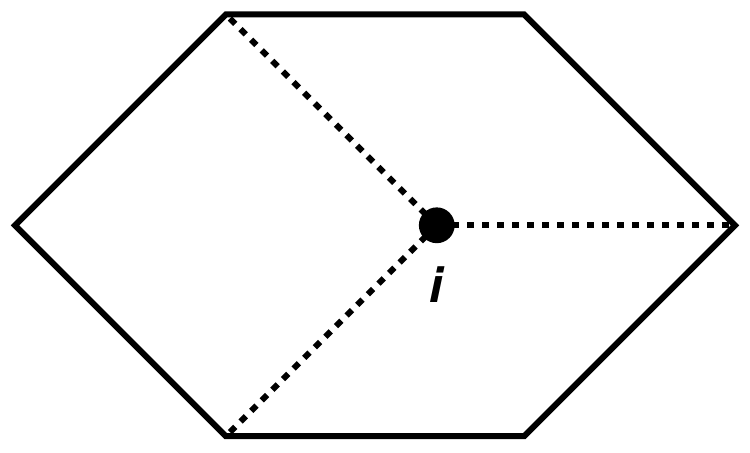}
 \hskip2cm \includegraphics[height=80pt] {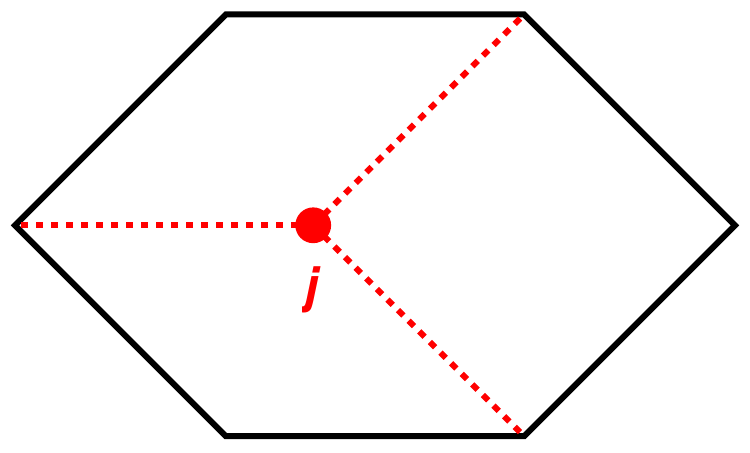}
\caption{A phason flip : the central site marked $i$ (left-hand figure) disappears, and is replaced by a new site $j$ (right hand figure). The black dashed bonds (left) are replaced by dashed red bonds (right). The 6 surrounding sites forming the outer hexagonal ring are not affected by the phason-flip, which results in a local permutation of a square and the two rhombuses.}
\label{fig:phasonflips}
\end{figure}

Shifting the position of the selection window in perpendicular space produces tilings which are equivalent in the sense of being locally isomorphic. \footnote{That is, two different tilings produced by different windows can be brought into coincidence in an arbitrarily large region by a suitable relative shift.} One can ask, however, how a particular tiling is changed when its selection window is shifted infinitesimally. The shift results in new sites appearing while some others disappear. This is illustrated for a small region in Fig.\ref{fig:phasonflips}). The original tiling, shown on the left, has a site labeled $i$ which becomes unselected and disappears, while the site labeled $j$ appears in the new tiling, shown on the right. The three dashed black bonds are replaced by the three red dashed bonds. All the surrounding sites remain unaffected. The result is a local rearrangement of tiles within the hexagonal region shown, called a ``phason-flip" .  The term phason excitation refers to the long range phenomenon of $correlated$ phason flips $throughout$ the quasicrystal that are associated with an infinitesimal window shift in perpendicular space. We will show an example of such correlated phason flips for a periodic approximant in the next section (see Fig.\ref{fig:rowphasonflips}). 

\subsection{Random phason flips and randomized tilings.}
We have focused thus far on how to generate perfect AB tilings, which are members of a wider family of deterministic square-rhombus tilings. Perfect tilings are in turn members of the larger set of random tilings made from the same tiles. Random tilings have been proposed as models for real quasicrystals \cite{henley}, as they are favored by entropic considerations (there are a macroscopic number of random tilings possible for a given system size, whereas there are only a handful of perfect tilings). 

\begin{figure} [H] 
\centering
\includegraphics[height=150pt] {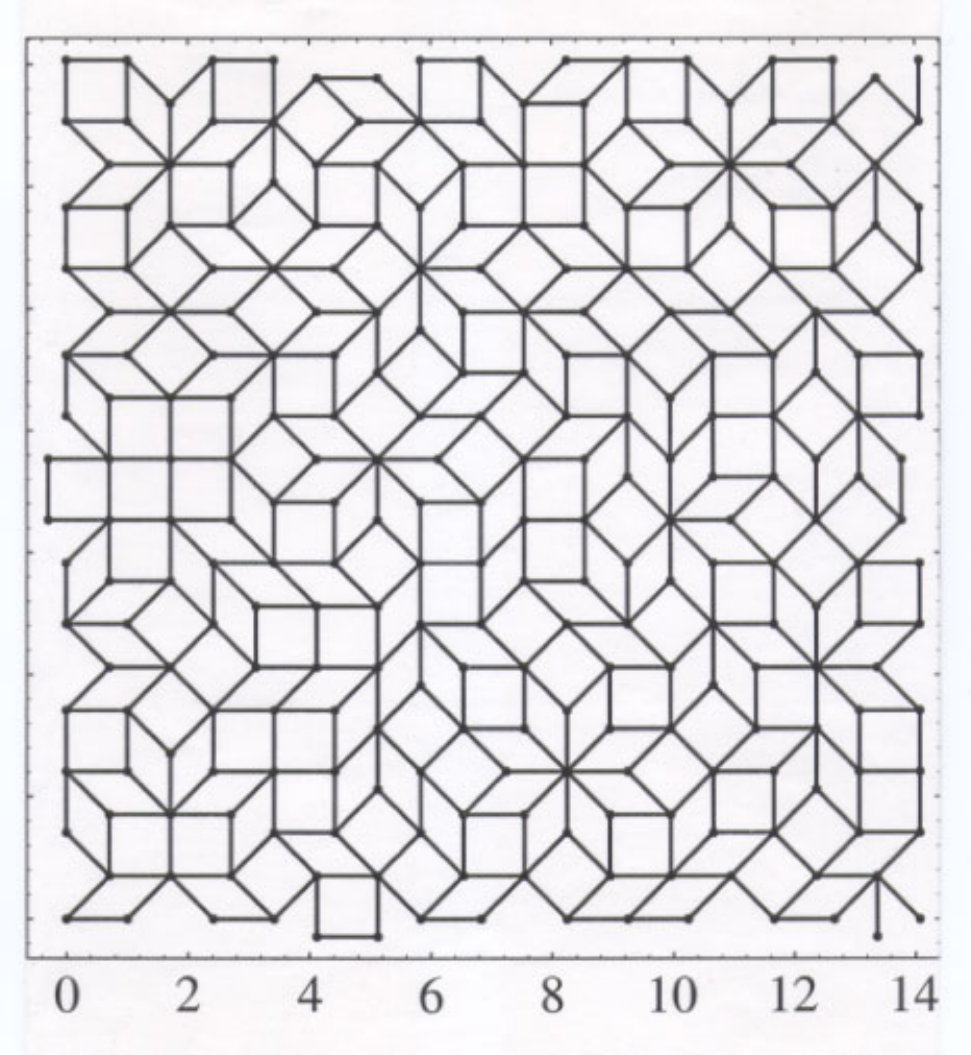}
\caption{A randomized approximant tiling obtained by successive independent phason flip operations carried out on a $N=239$ site approximant.}
\label{fig:randomtiling}
\end{figure}

One is led to ask how properties of random tilings compare with those of perfect tilings. To study properties of random tilings, it is possible to randomize the perfect tilings described above, to varying degrees by performing so-called local ``phason flips". These local operations will be defined now. We have seen in the previous sub-section that infinitesimally shifting the selection window $W$ results in a collective phason mode, i.e. in a correlated set of displacements across the entire tiling. One such collective mode is shown in Fig.\ref{fig:rowphasonflips} for an approximant tiling. This phason  can be regarded as a set of simultaneous local flips occurring along a line. Each flip is of the type illustrated in Fig.\ref{fig:phasonflips}. The central site is $F$-type.
To obtain a randomized version of the perfect tiling, one can thus proceed as follows: 
\begin{enumerate}
\item identify all possible flip sites ($F$ sites of the tiling) and randomly select one 
\item do a phason flip and redraw the bonds at the chosen site
\item repeat the above steps 
\end{enumerate}
After a number of repetitions, a randomized tiling such as the one in Fig.\ref{fig:randomtiling} is obtained. Note that this tiling has local environments which are not present in the perfect AB tiling, such as a vertex surrounded by three squares. It can be shown that the structure factor of random tilings has a continuous background and the Bragg peaks of the perfect tiling are broadened \cite{henley}. The real space repetitivity property (Conway theorem) is lost in these tilings. Random tilings do not have strict inflation-deflation symmetry. The loss of these real space properties has many consequences for electronic properties, including spectral statistics \cite{jagapiech,grimmroemer}, wave function decay power laws \cite{tarzia}, and quantum diffusion of wave packets and transport \cite{trambly,tarzia}. 

\section{Periodic approximants of the Ammann-Beenker tiling}
The perfect AB tiling of the previous section has been obtained by projecting onto an irrational plane $E$.  If this irrational plane is slightly tilted, to a rational orientation, the resulting tiling is a periodic structure. This can be done using the approximants of the irrational number $\lambda$ which underlies the AB tiling.  We introduce now the matrices
$M_n = \Pell_n M + \Pell_{n-1}I$, where $\Pell_n$ are the Pell numbers (see Appendix). Explicitly, one has
\begin{align}
\label{Mnmatrix}
M_n = \begin{pmatrix}
\Pell_n+\Pell_{n-1} & \Pell_n & 0 & -\Pell_n  \\
\Pell_n & \Pell_n+\Pell_{n-1} &  \Pell_n &  0  \\
0 &  \Pell_n & \Pell_n+\Pell_{n-1} & \Pell_n  \\
- \Pell_n &  0 & \Pell_n &  \Pell_n+\Pell_{n-1} \\
\end{pmatrix} .
\end{align}

The inverse of the inflation matrix $M^{-n}$ is given by
\begin{align*}
M^{-n} &= (-1)^n \left( -\Pell_n M + \Pell_{n+1} I \right) \\
&= (-1)^n \begin{pmatrix}
\Pell_n+\Pell_{n-1} & -\Pell_n & 0 & \Pell_n  \\
-\Pell_n & \Pell_n+\Pell_{n-1} &  -\Pell_n &  0  \\
0 &  -\Pell_n &\Pell_n+\Pell_{n-1} & -\Pell_n  \\
\Pell_n &  0 & -\Pell_n & \Pell_n+\Pell_{n-1} \\
\end{pmatrix} .
\end{align*}
The series of approximants is obtained by considering a series of rational planes $E_n$ that will converge in the limit of infinite $n$
to $E$. These planes are obtained from an initial plane by iteration using $M$.

Projection matrices for the $n$th approximant are defined, in analogy with those defined in the previous section for the quasiperiodic tiling, as follows
\begin{align}
\label{projectors2dappxt}
\pi_n = \frac{1}{\mathcal{N}_n}\begin{pmatrix}
\Pell_n+\Pell_{n-1} & \Pell_n & 0 & -\Pell_n  \\
 0 & \Pell_n & \Pell_n+\Pell_{n-1} & \Pell_n 
\end{pmatrix} , \nonumber \\
\pi'_n = \frac{1}{\mathcal{N'}_n}\begin{pmatrix}
2 \Pell_n & -(\Pell_n+\Pell_{n-1}) & 0 & \Pell_n+\Pell_{n-1} \\
0 & \Pell_n+\Pell_{n-1} & -2 \Pell_n & \Pell_n+\Pell_{n-1}   \\
\end{pmatrix} .
\end{align}
where the normalization factors are given by $2\mathcal{N}_n^2=4\Pell_n^2 + (-1)^n$ and ${\mathcal{N}'}_n^2=2\mathcal{N}_n^2$.

A basis set of the rational plane $E_n$ can be defined as follows
\begin{align*}
\vec{e}_{x,n} &=  \frac{1}{\sqrt{2}\mathcal{N}_n} \begin{pmatrix}
\Pell_{n}+\Pell_{n-1}   \\
\Pell_{n}   \\
0  \\
-\Pell_{n} \\
\end{pmatrix} \nonumber \\
\vec{e}_{y,n} &=  \frac{1}{\sqrt{2} \mathcal{N'}_n} \begin{pmatrix}
0  \\
\mathcal{N}_n   \\
\mathcal{N}_n+\Pell_{n-1}   \\
\mathcal{N}_n \\
\end{pmatrix} 
\end{align*}
in the basis $\{e_0,e_2\}$ and a basis set of the rational plane $E'_n$ can be defined as follows 
\begin{align*}
\vec{e}'_{x,n} &=  \frac{1}{2\mathcal{N}_n} \begin{pmatrix}
2\Pell_{n}   \\
-(\Pell_{n}+\Pell_{n-1})   \\
0  \\
\Pell_{n}+\Pell_{n-1} \\
\end{pmatrix} \nonumber \\
\vec{e}'_{y,n} &=  \frac{1}{2\mathcal{N'}_n} \begin{pmatrix}
0  \\
\Pell_{n}+\Pell_{n-1}  \\
-2 \Pell_{n}  \\
\Pell_{n}+\Pell_{n-1} \\
\end{pmatrix} 
\end{align*}
in the basis $\{e'_1,-e'_3\}$ (where the minus sign is added to make the basis direct).
To avoid ambiguities due to projections which lie on top of its boundaries, the window $W_n$ must be shifted away from the origin. The resulting approximant loses $D_4$ symmetry (of the square, with 4 mirror planes) in the strict sense. However they have an approximate $D_4$ symmetry. If one overlays an approximant and the tiling obtained by a $\pi/2$ rotation, they overlap perfectly, except along two orthogonal 1D stripes, also called ``worms" \cite{grimmreview}. The approximants also possess one exact reflection symmetry with respect to one of the diagonals. 

To illustrate the method, we now give some results for the $n=2$ approximant. The relevant Pell numbers are $\Pell_1=1$ and $\Pell_2=2$. The normalization constant $\mathcal{N}=\sqrt{17/2}$. The basis vectors of the plane $E_2$ are 
\begin{align}
\label{eq:basisappxt5}
\vec{e}_{2,x} = \sqrt{\frac{1}{17}}\begin{pmatrix}
3 \\
2 \\
0\\
-2
\end{pmatrix},
~ \vec{e}_{2,y} = \sqrt{\frac{1}{17}}\begin{pmatrix}
0 \\
2 \\
3\\
2
\end{pmatrix}
\end{align}
The projection matrix onto the plane $E_2$ is 
\begin{align}
\label{eq:appxt5}
\pi_2 = \sqrt{\frac{1}{17}}\begin{pmatrix}
3 & 2 & 0 & -2  \\
 0 & 2 & 3 & 2 
\end{pmatrix} 
\end{align}

A point $\vec{r}= x \vec{e}_{2,x} + y \vec{e}_{2,y}$ lying in $E_2$ has the coordinates $\{R_1,R_2,R_3,R_4\}$ in $\ZZ_4$ given by
\begin{align}
\label{eq:z4appxt2}
\begin{pmatrix}
R_1 \\
R_2 \\
R_3\\
R_4
\end{pmatrix}
= \sqrt{\frac{1}{17}}\begin{pmatrix}
3 & 0 \\
2 & 2 \\
0 & 3\\
-2 & 2
\end{pmatrix} \begin{pmatrix}
x \\
y
\end{pmatrix}
\end{align}
In the perpendicular plane, we have $\mathcal{N}'=\sqrt{17}$. The basis vectors of $E'_2$ are given by 
\begin{align}
\label{eq:basisperpappxt5}
\vec{e}'_{2,x} = \sqrt{\frac{1}{17}}\begin{pmatrix}
4 \\
-3 \\
0\\
3
\end{pmatrix},
~ \vec{e}'_{2,y} = \sqrt{\frac{1}{17}}\begin{pmatrix}
0 \\
3 \\
-4\\
3
\end{pmatrix}
\end{align}
The projection matrix in perpendicular space is
\begin{align}
\label{eq:perpappxt5}
\pi'_2 = \sqrt{\frac{1}{17}}\begin{pmatrix}
4 & -3 & 0 & 3  \\
 0 & 3 & -4 & 3 
\end{pmatrix} 
\end{align}
The window of the approximant $n=2$ and the approximant are shown in Fig.\ref{W_3}, with site indices to show corresponding positions. 
\begin{figure} [H] 
\centering
\includegraphics[height=200pt] {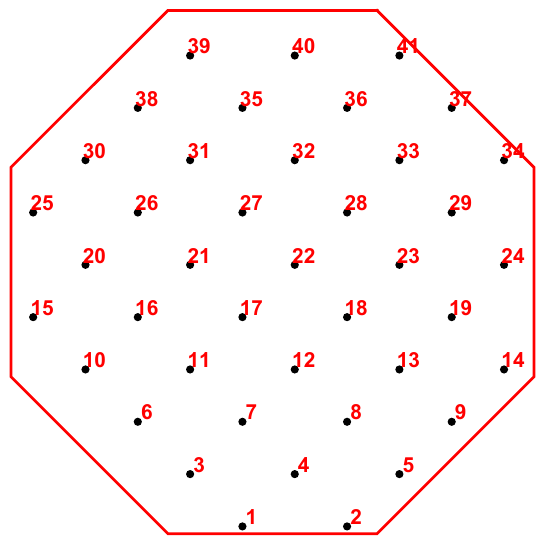} \includegraphics[height=200pt] {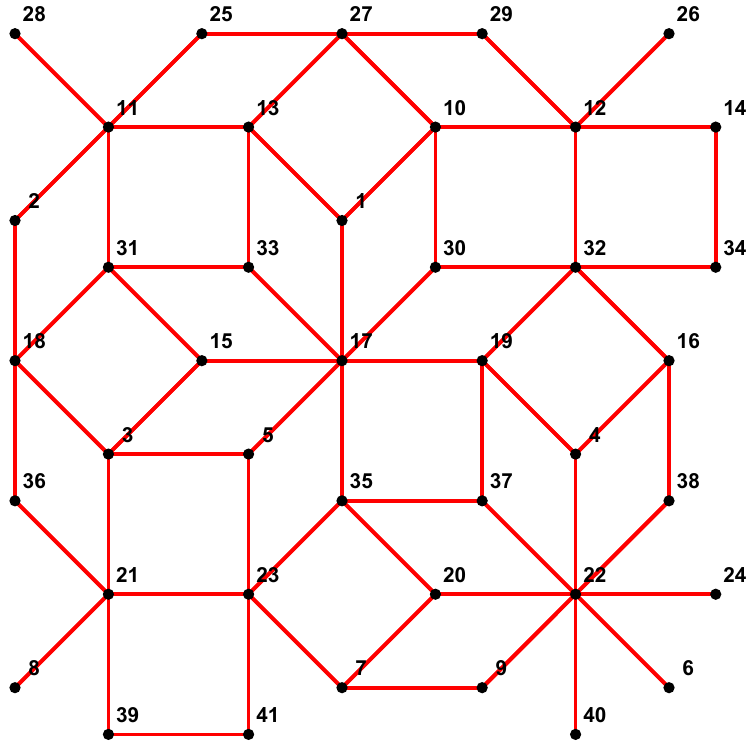}
\caption{The selection window $W_2$ (left) and the projected approximant tiling (right) are shown for $n=2$. The vertices have been indexed to show the correspondence between parallel space and perpendicular space positions.}
\label{W_3}
\end{figure} 
It can be noted that the tiles in periodic approximants generated by this method are slightly deformed in terms of angles and edge lengths. In Appendix B we describe an alternative method by which square approximants with perfect squares and 45$^\circ$ rhombuses may be obtained.

\subsection{Frequencies of vertex types}
The octagon $W_n$ is subdivided into domains labeled $A,B,C,D,E,F$ corresponding, as already seen, to the six different local enviroments of the tiling in $E$. Sites of the family $F$ (coordination number 3) correspond to the eight triangular domains $F_n$, whose total surface area is 
$S(F_n) = 2\Pell_n^2 +(\Pell_n+\Pell_{n-1})^2 = Q_{2n+1}$ (see Appendix). 

The areas subtended by the domains corresponding to each type of vertex are indicated below.

\begin{table}[H]
\begin{center}
\small
\begin{tabular}{|c|c|r|r|r|r|}
\hline
Domains & Surfaces & $n=2$ & $n=3$ & $n=4$ & $n=5$ \\
\hline
A & $ Q_{2n-3} $ & 1 & 7 & 41 & 239\\
B & $ Q_{2n-4} $ & 1 & 3 & 17 & 99\\
C &  $ 2 Q_{2n-3} $ & 2 & 14 & 82 & 478\\
D & $ 2 Q_{2n-2} $ &  6 & 34 & 198 & 1154\\
E & $ 2 Q_{2n-1} $ & 14 & 82 & 478 & 2786\\
F & $ Q_{2n}$ & 17 & 99  & 577 & 3363\\
\hline
Sum & $Q_{2n+1}$ & 41 & 239 & 1393 & 8119\\
\hline
\end{tabular}
\caption{Areas of subdomains $A$, $B$, $C$, $D$, $E$ and $F$ for the $n$th approximant in $W_n$.}
\label{Surfaces}
\normalsize
\end{center}
\end{table}
The areas can be more simply expressed in terms of $Q_n=\Pell_n+\Pell_{n-1}$, which satisfy the  same recursion relation as the Pell numbers $Q_{n+1}=2Q_{n}+Q_{n-1}$ but are given by initial conditions  $Q_0=1$, 
$Q_1=1$. The total number of sites in the $n$th approximant is then given by $Q_{2n+1}$, while the sub-domains have areas given by
\begin{eqnarray}
\{S_A,....,S_F\} = \{  Q_{2n-3}, Q_{2n-4},  2Q_{2n-3}, Q_{2n-2}, 2Q_{2n-1},   Q_{2n}     \}
\end{eqnarray}

\begin{figure} [H] 
\centering
\includegraphics[height=140pt] {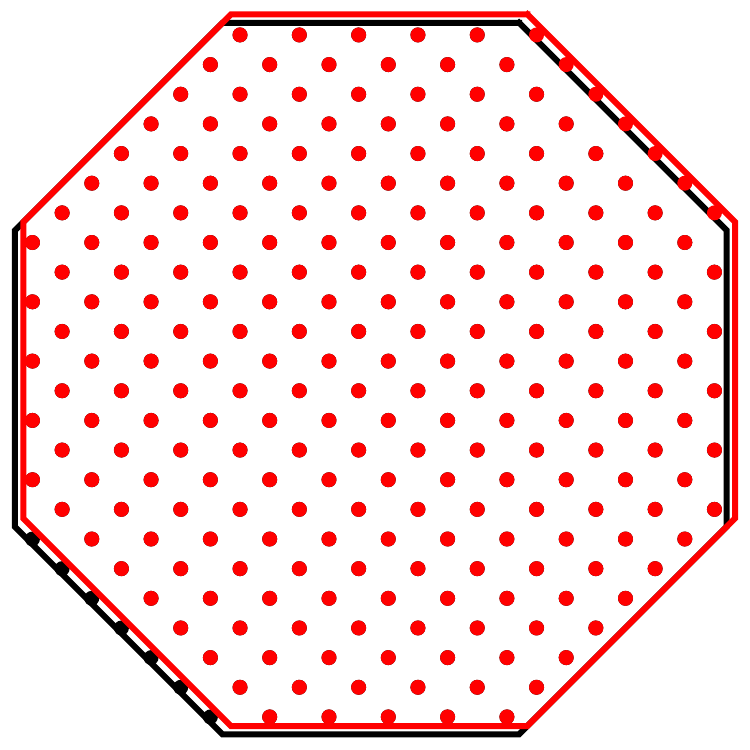} \hskip1cm \includegraphics[height=140pt] {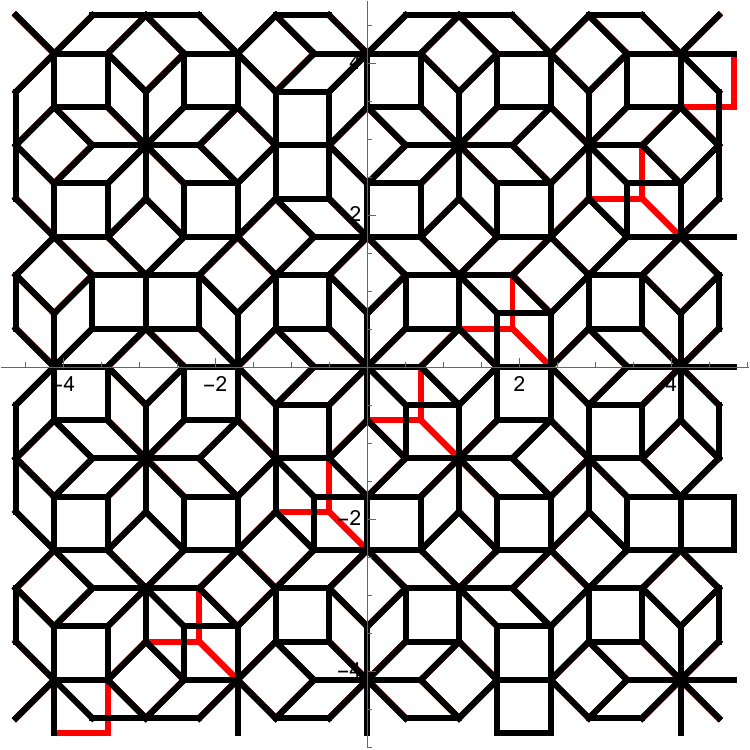}
\caption{(Left) A small diagonal shift in perpendicular space of the window results in some old points (in black) being unselected and new (in red)  points being selected. (Right) The resulting new and old tilings superpose everywhere except for phason flips along a single row parallel to the diagonal.}
\label{fig:rowphasonflips}
\end{figure}

As mentioned in the previous section, a row of phason flips can be generated by a small displacement of the selection window. This is illustrated in Figs.\ref{fig:rowphasonflips}. The left-hand figure shows the selected points before (gray) and after (red) a small shift along a diagonal direction of the window. The new window unselects a row of old (gray) points and replaces these by new (red) points. The right-hand figure shows that the resulting tilings agree everywhere except along a row parallel to the diagonal, where phason-flips appear.

\subsection{Matching rule defects}
\begin{figure} [H] 
\centering
\includegraphics[height=250pt] {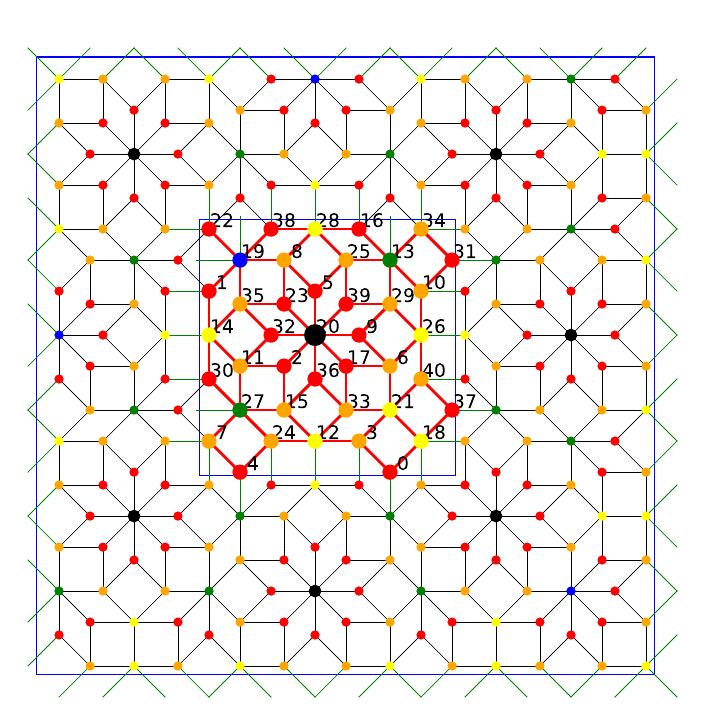}
\caption{Superposition of two approximants ($n=2$ and $n=3$), with the unit cell of approximant $n=2$ indicated by thick red edges. One sees that there is a perfect overlap of all of the sites corresponding to the unit cell of the smaller approximant. However upon applying periodic boundary conditions, one sees that there results a mismatch of bonds (four defects, along the top and left hand edges).  }
\label{fig:overlap}
\end{figure}

Since approximants periodically repeat in the plane, their structure differs from that of the quasiperiodic tiling, when regions of large enough size are compared. An important question concerns the number and the nature of these discrepancies or defects in approximants. Zijlstra \cite{zijlstra} observed that the smallest approximant violates matching rules at the boundary. Using inflation rules, he argued that the size of this defected region must increase linearly with system size, that is, more slowly than the total number of sites. Another way to visualize these defects is illustrated in Fig.\ref{fig:overlap}, which shows the superposition of a smaller ($n=2$) approximant within the next biggest ($n=3$) approximant. One can see that all the sites exactly coincide within the unit cell of the smaller tiling. At the unit cell boundary, however, one sees that defects occur. Some sites of the larger tiling do not coincide with the periodic continuation of the smaller tiling. In the example shown in the figure, the mismatches occur at the sites labeled 31 and 37, as well as sites 0 and 4 (Nb. labels are different in this figure as compared to Fig.\ref{W_3}. It can be checked similarly, that the $n=3$ approximant overlaps with the $n=4$ approximant in its interior but that there are discrepancies at the boundary where bonds do not match. The number of defects grows linearly with the edge length of the approximant, as noted by Zijlstra.

The presence of such defects is expected to have consequences for physical properties computed using periodic approximants. However, depending on the nature of quantities which are computed, their effects may be more or less important. The so-called ``height-field" which enters in the construction of the ground state wave-function of tight-binding models \cite{mace2017} is not single-valued in these tilings. It is in this case preferable to work with patches with ``mirror" boundary conditions as introduced by Kalugin and Katz \cite{kalugin2014}. Kalugin and Katz showed that the ground state energy computed using the latter converges faster compared to using square approximants. On the other hand, when short range correlations are the main determining factor, as for quantities which are integrated over occupied states (the mean field order parameters for magnetic phases, for example), the deviations from ``pure" quasiperiodic systems should be very small, if sufficiently large approximants are considered. Finally, the defects are topological perturbations which may play a role, for example, in the computations of topological invariants. They may also be relevant to consider for problems where the tiling is subjected to a magnetic flux. These questions remain to be addressed in future studies. 

\section{Conclusions} Many interesting physical properties of quasicrystals can be studied using simplified models defined on quasiperiodic tilings and their periodic approximants. The AB tiling is interesting not only because there exist real quasicrystals with eight-fold symmetry, but because it is a good paradigm for 2D structures having long range quasiperiodic order. As pointed out in the introduction, this tiling is often chosen for reasons of simplicity of its geometry. We note that an even simpler variant of this tiling, called the labyrinth, has been introduced by Sire et al \cite{siremossadoc}. It can be constructed from a direct product of 1D silver mean chains, and all vertices have coordination number 4. Its structure allows for some important simplifications compared to the AB tiling. One has notably, a separable Hamiltonian for the labyrinth which writes as a sum of two 1D Hamiltonians \cite{thiemschr}. This property does not extend to the AB tiling which is so to say a ``genuinely" two-dimensional quasiperiodic system. 

In this brief review the Cut-and-Project method to obtain the Ammann-Beenker tiling is introduced, and some of its principal geometric properties are described. The procedures to obtain periodic approximants are described and many of their properties are outlined. Throughout the emphasis is placed on practical issues, as this is intended to be a working guide. For derivations and more details of the CP method, the reader is referred to the original papers cited in the text. 

More complex variants of this tiling can be found in the literature -- for example, Godreche et al have introduced square-rhombus tilings having 8-fold symmetry, but more complicated selection windows, which can be fractal \cite{godreche}. These tilings are characterized by diffraction patterns which are not point-like. It would be most interesting to study their electronic properties, which can be expected to differ quite significantly from those of the AB tiling. 

In a different context, we have shown that it is possible to obtain the standard AB tiling by trapping atoms in an optical potential made by superposition of four laser beams \cite{jaga2013}. These optical quasi-lattice structures can also be described by a cut-and-project scheme. Their most important advantage is they are (in principle) easily generated in an experimental setup. However -- showing there is no such thing as a free lunch -- they are intrinsically defected with respect to the perfect tiling -- a fraction of the sites are pairs related by phason flips (and not simultaneously occupied in the perfect tiling) along with a fraction of empty sites.

Some calculations and figures were made with SageMath, the Sage Mathematics Software System (Version 10.0), 
The Sage Developers, 2023, \url{https://www.sagemath.org}.

\appendix
\section{The Pell numbers}
The series of Pell numbers  $\{\Pell_n\}$ is defined for $n\geq 0$ by the initial conditions $\Pell_0=0$, 
$\Pell_1=1$, and the recursion relation $\Pell_{n+1}=2\Pell_{n}+\Pell_{n-1}$ for $n\geq1$. Thus
\begin{align}
\{ \Pell_n \}  = \{ 0, 1, 2, 5,12, 29, 70,169,408,985,2378,... \} .
\end{align} 
One has the following matrix relation 
\begin{align}
\label{Omatrix}
\begin{pmatrix}
\Pell_{n+1}\\
\Pell_n
\end{pmatrix}  
= T \begin{pmatrix}
\Pell_n\\
\Pell_{n-1}
\end{pmatrix} .
\end{align}
where $T$ is the symmetric matrix given by 
\begin{align}
\label{Tmatrix}
T &= \begin{pmatrix}
\Pell_2 & \Pell_1 \\
\Pell_1 & \Pell_0
\end{pmatrix}  
= \begin{pmatrix}
2 & 1 \\
1 & 0
\end{pmatrix} . 
\end{align}
It is easily seen that the recursion relation implies that 
\begin{align}
\begin{split}
\label{Tn}
\begin{pmatrix}
\Pell_{n+1} & \Pell_n \\
\Pell_{n} & \Pell_{n-1}
\end{pmatrix} &= T
\begin{pmatrix}
\Pell_{n} & \Pell_{n-1} \\
\Pell_{n-1} & \Pell_{n-2}
\end{pmatrix}
= T^{n-1} \begin{pmatrix}
\Pell_2 & \Pell_1 \\
\Pell_1 & \Pell_0
\end{pmatrix} = T^n.
\end{split}
\end{align}
From this it can be deduced that 
\begin{align}
\det T^n = \Pell_{n+1}\Pell_{n-1}-{\Pell_n}^2 = (-1)^n .
\end{align}
The relation (\ref{Tn}) implies that $T^{p+q} = T^p T^q$ :
\begin{align}
\begin{split}
\label{Tpq}
\begin{pmatrix}
\Pell_{p+q+1} & \Pell_{p+q} \\
\Pell_{p+q} & \Pell_{p+q-1}
\end{pmatrix} &= 
\begin{pmatrix}
\Pell_{p+1} & \Pell_p \\
\Pell_p & \Pell_{p-1}
\end{pmatrix}
\begin{pmatrix}
\Pell_{q+1} & \Pell_q \\
\Pell_q & \Pell_{q-1}
\end{pmatrix}
\end{split}
\end{align}
In particular one has the following relations :
\begin{align}
\begin{split}
\Pell_{2n} &=  \Pell_{n-1}\Pell_n + \Pell_n\Pell_{n+1}, \\
\Pell_{2n+1} &=  \Pell_{n-1}\Pell_{n+1} + \Pell_n\Pell_{n+2} =  \Pell_n^2 + \Pell_{n+1}^2 . \\
\end{split} 
\end{align}

The characteristic polynomial of $T$ is $\det(T-xI) = x^2-2x-1$ and the two eigenvalues are the ``silver mean" 
$\lambda = 1+\sqrt{2}$ and its conjugate $\lambda' = 1-\sqrt{2} = -\lambda^{-1}$. $T$ satisfies the equation $T^2 = 2T+I$ and as a result one finds that $T^n = \Pell_nT + \Pell_{n-1}I$ for $n\geq 2$. 
One thus has 
\begin{align*}
T\begin{pmatrix}
\lambda \\
1
\end{pmatrix} = 
\lambda \begin{pmatrix}
\lambda \\
1
\end{pmatrix}, \quad
T^n.
\begin{pmatrix}
\lambda \\
1
\end{pmatrix} 
= \begin{pmatrix}
\lambda^{n+1} \\
\lambda^{n}
\end{pmatrix} ,
\end{align*}
and the same relation holds for $\lambda'$. Using (\ref{Tn}) one obtains the relation 
\begin{align}
\begin{split}
\lambda^{n} &= 
(0 \,,1)
T^n 
\begin{pmatrix}
\lambda \\
1
\end{pmatrix} =
(\Pell_{n}\,, \Pell_{n-1} ) 
\begin{pmatrix}
\lambda \\
1
\end{pmatrix} 
= \lambda \Pell_{n}  + \Pell_{n-1} 
\end{split} ,
\end{align}
satisfied by both $\lambda$ and $-\lambda^{-1}$. For $n\geq 1$ on also finds :
\begin{align}
\label{tau^ntau^-n}
(-\lambda)^{-n} &= -\lambda^{-1} \Pell_n + \Pell_{n-1} = -\lambda \Pell_n + \Pell_{n+1} .
\end{align}
From the second equation one deduces the limiting value of $\Pell_{n+1}/\Pell_n$ : 
\begin{align*}
\frac{\Pell_{n+1}}{\Pell_n} - \lambda = \frac{(-\lambda)^{-n}}{\Pell_n} \Rightarrow
\lim_{n\to \infty} \frac{\Pell_{n+1}}{\Pell_n} = \lambda.
\end{align*}
Upon subtracting the two relations in (\ref{tau^ntau^-n}) the following relations are obtained :
\begin{align}
\begin{split}
\lambda^{n} + (-1)^{n-1} \lambda^{-(n)} &= 2\lambda \Pell_{n}+\Pell_{n-1} - \Pell_{n+1} = 2(\lambda -1)\Pell_{n} , \\
\Pell_{n} &= \frac{\lambda^{n} + (-1)^{n-1} \lambda^{-n} }{2(\lambda -1)} 
= \frac{\lambda^{n} + (-1)^{n-1} \lambda^{-n} }{2\sqrt{2}} .
\end{split} 
\end{align}
The last equation allows for a rapid computation of $\Pell_n$ using the function $round$. \\
Upon adding the two equations in (\ref{tau^ntau^-n}) one finds that 
\begin{align}
\lambda^n + (-1)^n \lambda^{-n} = \Pell_{n-1} + \Pell_{n+1},
\end{align}
so that the integer closest to $\lambda^n$ is $\Pell_{n-1} + \Pell_{n+1}$ and the fractional  difference is 
\begin{align}
\lambda^n - \Pell_{n-1} - \Pell_{n+1} = (-1)^{n+1} \lambda^{-n} . 
\end{align} 
The quotients $\Pell_{n+1}\Pell_n$ are the convergents of the continued fraction of $1+\sqrt{2}$ which is  $[2;2,2,...]$. \\

The series $Q_n = \Pell_n + \Pell_{n-1}$ satisfies the same recursion formula as the Pell numbers but with initial conditions 
$Q_0 = Q_1 = 1$. Then 
\begin{align}
\begin{split}
Q_{2n} &=  \Pell_{2n} + \Pell_{2n-1} 
= \Pell_{n-1}\Pell_n + \Pell_n\Pell_{n+1} + \Pell_{n-1}^2 + \Pell_{n}^2 \\
&= 3\Pell_n^2 + 2\Pell_n\Pell_{n-1} +  \Pell_{n-1}^2, \\
Q_{2n-1} &= \Pell_{2n-1} + \Pell_{2n-2} = \Pell_n^2 + \Pell_{n-1}^2 
+ \Pell_{n-2}\Pell_{n-1} + \Pell_{n-1}\Pell_{n} \\
&= \Pell_n^2 + 2 \Pell_n \Pell_{n-1} - \Pell_{n-1}^2.
\end{split} 
\end{align}
Quotients $Q_{n+1/}Q_n$ are the convergents of the continued fraction of $2 + \sqrt{2}$ which is  $[3;2,2,...]$. \\

\section{Square rational approximants: An alternative method}

\begin{figure} [H] 
\centering
\includegraphics[height=130pt] {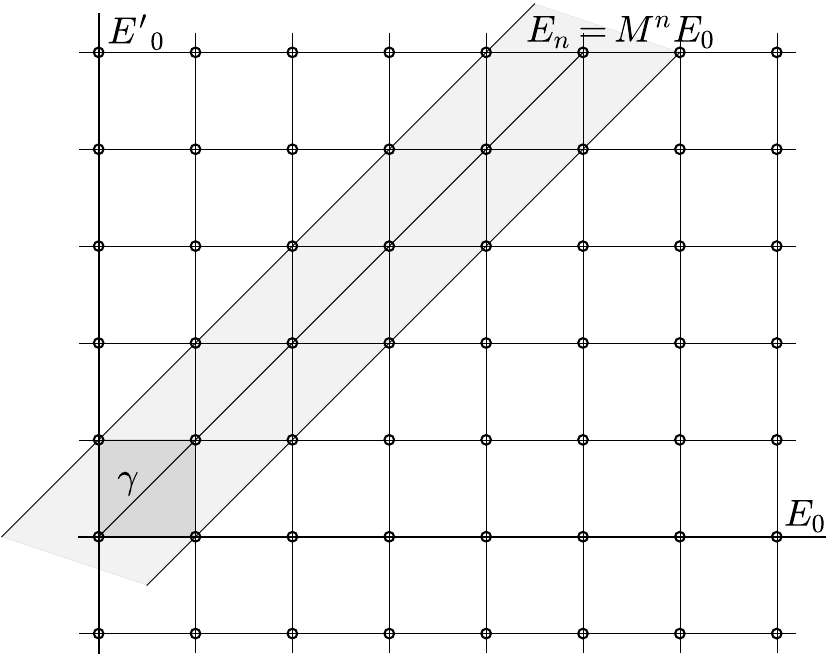}
\includegraphics[height=130pt] {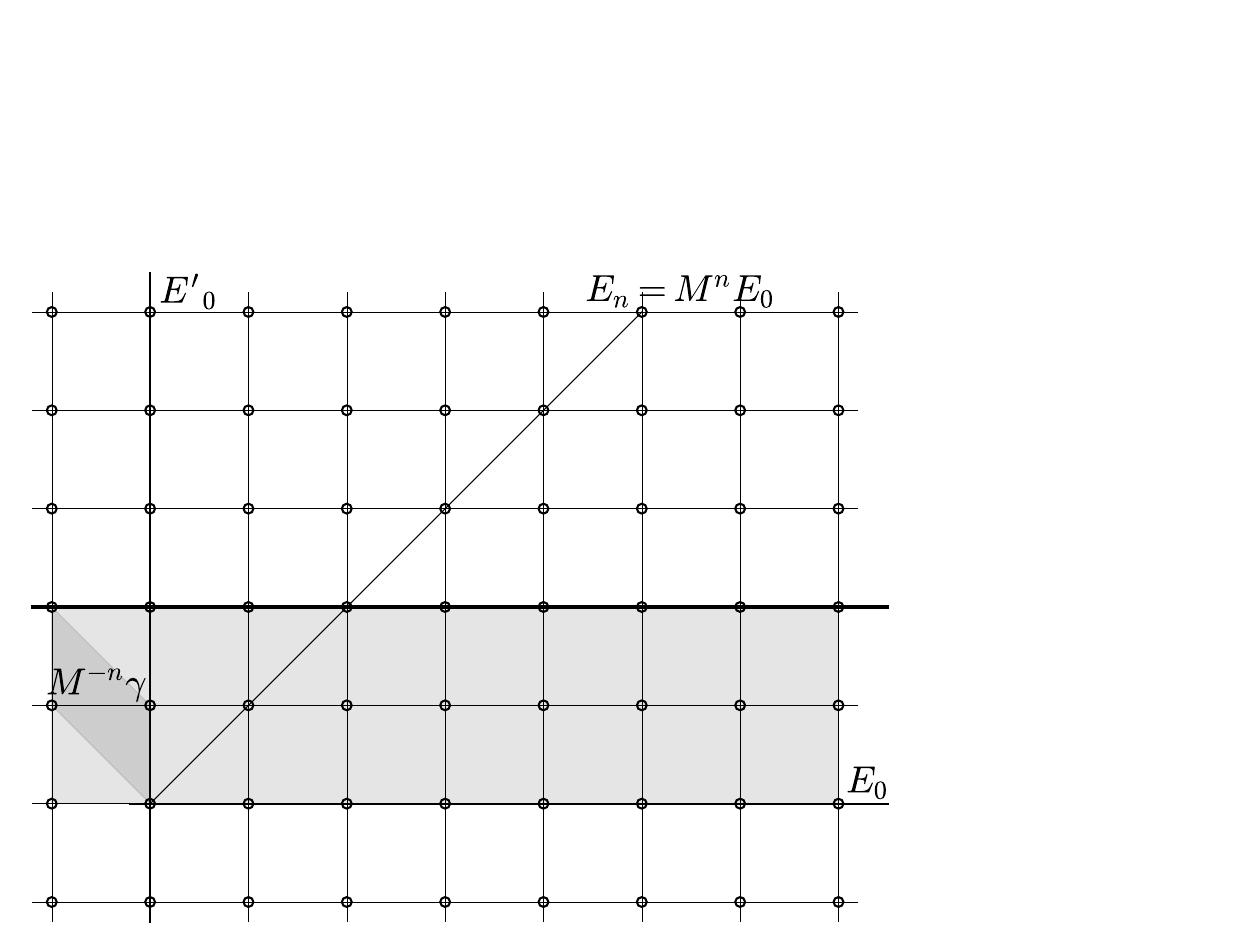}
\caption{Alternative CP method for approximants: (Left) The initial selection method. (Right) Transformation by $M^{-n}$ with constant parallel plane $E_0$ and constant perpendicular plane $E'_0$.}
\label{fig:cpe}
\end{figure}

An alternative method to handle rational approximants, first presented in \cite{duneau89}, is briefly recalled before application to square approximants of the quasiperiodic Ammann-Beenker tiling. 
The general idea is to transform the original construction by a linear map so that the new 
"parallel" and "perpendicular" spaces are coordinate vector spaces (See Fig.\ref{fig:cpe}). The only element depending on the order $n$ of the approximant is the selection window, a subset $W_n$ of the constant perpendicular space.

Let $E_0$ and $E'_0$ be the planes generated by $\{\eps_0,\eps_2\}$ and $\{\eps_1,\eps_3\}$ respectively. 
For any integer $n>0$ let $E_n=M^n E_0$ denote the plane spanned by the basis  
$\{M^n\eps_0,M^n\eps_2\}$ and let $\Lambda_n=\ZZ^4\cap E_n$.
The $n$th approximant is the projection on $E$ of a set of points $X_n$, belonging to the hypercubic lattice and lying close to $E_n$. If $\gamma$ denotes the unit cube, then $X_n = (E_n + \gamma) \cap \ZZ^4$ and  
the vertices of the tiling are the projection $p(\Xi_n)$ onto the plane $E$.{\footnote{In order to avoid singular 
cases where lattice points are on the boundary of $\Xi_n$ one considers sets $Xi_n = (E_n + \gamma +\tau)\cap \ZZ^4$ where $\tau$ is a translation.}} 

Applying the linear transformation $M^{-n}$ on the above construction we get the constant "parallel" space $E_0$ and the constant "perpendicular" space $E'_0$. The set $X_n$ is mapped to $M^{-n}X_n= (E_0+M^{-n}\gamma)\cap\ZZ^4$ and the selection window $W_n$ is the projection of $M^{-n}\gamma$ onto $E'_0$. A unit cell of $M^{-n}X_n$ is the set of lattice points 
$\ZZ^4\cap W_n$, lying in $E'_0$, and the set of selected points is given by all translates of this unit cell by the translation lattice $\Lambda_0 = \ZZ^4\cap E_0$ spanned by $\{{\eps}_0,{\eps}_2\}$. The perpendicular projection is particularly simple as it writes $(x_0,x_1,x_2,x_3) \to (x_1,x_3)$ and the vertices of $W_n$ are easily deduced from table \ref{octogone_index} and 
Fig.\ref{W_3}.

It remains to perform the inverse linear transformation $M^n$ on the unit cell and the translation lattice, followed by orthogonal projection onto the plane $E$, with $pM^n = \lambda^n p$. However the points are scattered on different unit cells of the approximant so that they must be gathered to get a connected unit cell by using a {\it frac} or {\it round} function.  This finally gives a unit cell of the $n$th approximant tiling and its translation lattice $p(\Lambda_n) = \lambda^n p(\Lambda_0)$.  

\subsection{Selection window for square approximants}
We give in the table below the 16 vertices obtained by projection onto the plane $E'_0$ generated by $\{{\eps}_1,{\eps}_3\}$ 
 of the 16 vertices of $\gamma_n$ in the basis $\{{\eps}_1,{\eps}_3\}$. 
\begin{table}[h]
\begin{center}
\small
\begin{tabular}{|c|c|c|c|c|c|}
\hline
id & vertex & coordinates & id & vertex & coordinates \\
\hline
0 & $(0,0,0,0)$ & $(0,0)$ & 8 & $(0,1,1,0)$ & $(\Pell_{n-1},-\Pell_n)$ \\
1 & $(1,0,0,0)$ & $(-\Pell_n,\Pell_n)$ & 9 & $(0,1,0,1)$ & $(\Pell_n+\Pell_{n-1},\Pell_n+\Pell_{n-1})$ \\
2 & $(0,1,0,0)$ & $(\Pell_n+\Pell_{n-1},0)$ & 10 & $(0,0,1,1 )$ & $(-\Pell_n,\Pell_{n-1})$ \\
3 & $(0,0,1,0)$ & $(-\Pell_n,-\Pell_n)$ & 11 & $(1,1,1,0 )$ & $(\Pell_{n-1}-\Pell_n,0)$ \\
4 & $(0,0,0,1)$ & $(0,\Pell_n+\Pell_{n-1})$ & 12 & $(1,1,0,1 )$ & $(\Pell_{n-1},\Pell_{n+1})$ \\
5 & $(1,1,0,0)$ & $(\Pell_{n-1},\Pell_n)$ & 13 & $(1,0,1,1 )$ & $ (-2\Pell_n,\Pell_n+\Pell_{n-1})$ \\
6 & $(1,0,1,0)$ & $(-2\Pell_n,0)$ & 14 & $(0,1,1,1 )$ & $(\Pell_{n-1},\Pell_{n-1})$ \\
7 & $(1,0,0,1)$ & $(-\Pell_n,\Pell_{n+1})$ & 15 & $(1,1,1,1)$ & $(\Pell_{n-1}-\Pell_n,\Pell_n+\Pell_{n-1})$ \\
\hline
\end{tabular}
\caption{Indices and coordinates of vertices of the 4D unit cube after transformation by $M^{-n}$ and projection on the 
plane $\{{\eps}_1,{\eps}_3\}$.}
\label{octogone_index}
\normalsize
\end{center}
\end{table} 
The windows for the approximant tilings are the octagonal domains $W_n$ whose total area is 
$S(W_n) = \Pell_{n+1}^2 + 2\Pell_{n+1}\Pell_n - \Pell_n^2 = Q_{2n+1}$. 
The 8-fold symmetry of the quasiperiodic tiling is lost. The windows $W_n$ can be thought of as the intersection of two squares of sides $a_n$ and $b_n$ where
\begin{align}
    a_n &= \Pell_{n+1} + \Pell_n \\ 
    b_n &= \sqrt{2} \Pell_{n+1} 
\end{align}
calculated from table \ref{octogone_index}. In the limit $n\rightarrow \infty$, $a_n/b_n$ tends to $1$, restoring 8-fold symmetry.

The centered window of the approximant $n=3$ is shown in Fig.\ref{W_3}. 
\begin{figure} [H] 
\centering
\includegraphics[height=220pt] {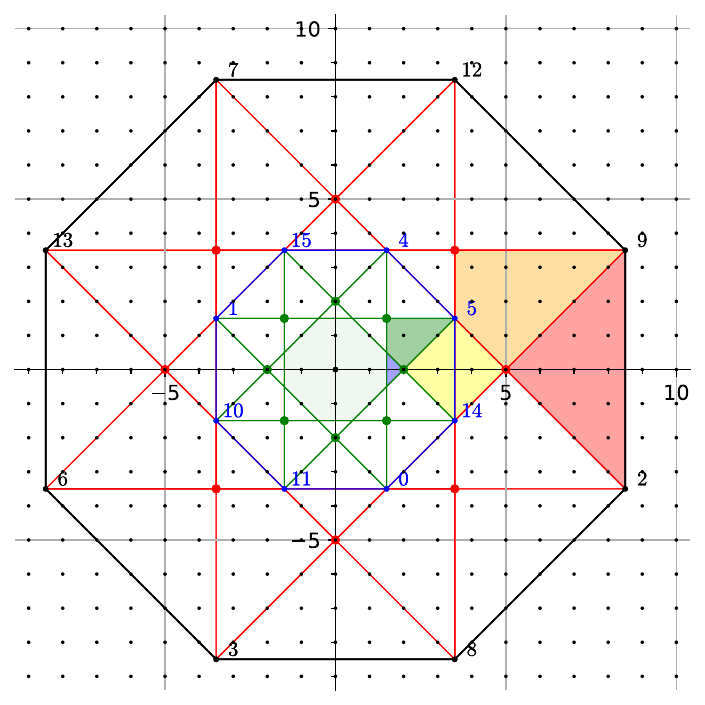}
\caption{The centered selection window $W_3$. The vertices of the unit cube  $\gamma_3 = M^{-3}\gamma$ have been indexed according to the scheme presented in table \ref{octogone_index}. $S(W_3) = 239$ and there are $99$ sites $F$. 
}
\label{W_3}
\end{figure} 

\begin{figure} [H] 
\centering
\includegraphics[height=160pt] {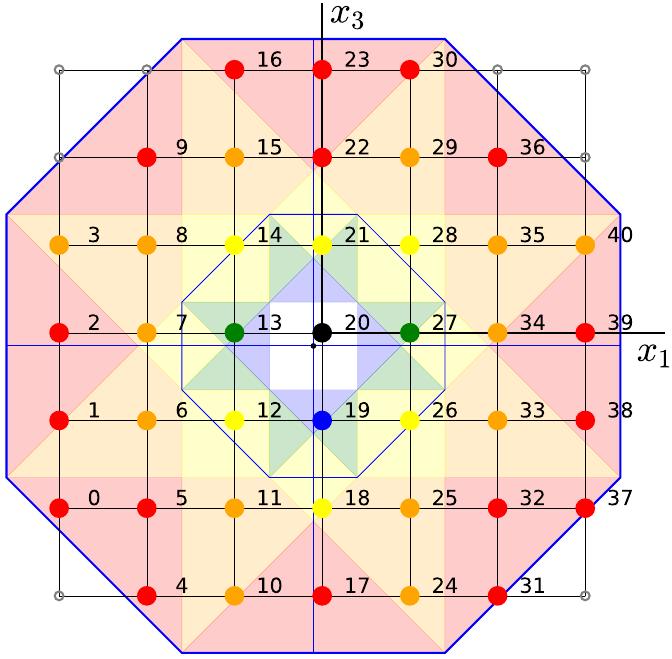}
\includegraphics[height=160pt] {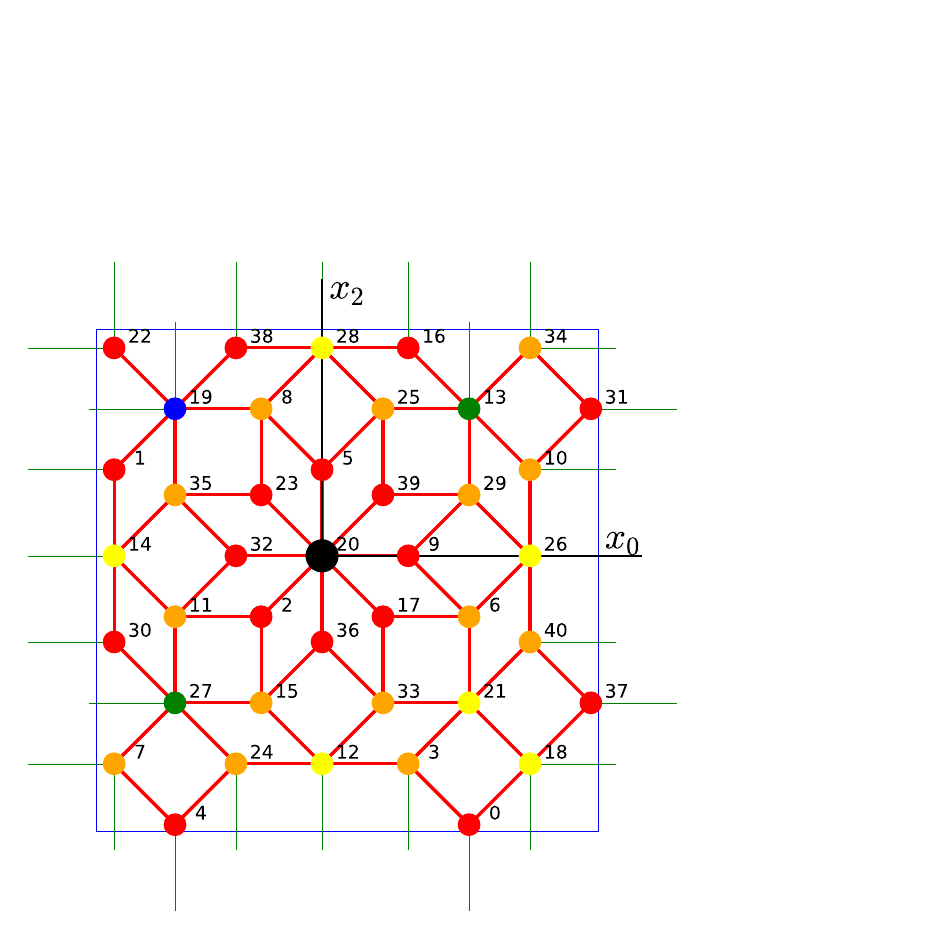}
\caption{(Left) Selection window in perpendicular space $E'_0$ for an approximant of order 2. Points are indexed by $i=0,..,40$. (Right) the corresponding projections on $E$ in basis $\{p\vec{\eps}_0,p\vec{\eps}_2\}$. 
}
\label{fig:cell2}
\end{figure}

Figs.\ref{fig:cell2} illustrates the method, for the case of $n=2$. The lefthand figure shows the region in perpendicular space $E'_0$ with the selection window $W_2$, and projections of selected points indexed by an integer $i$. The selection window has been shifted by the small displacement $(0.1,0.15)$. The right-hand figure shows the projections of the same points in parallel space, with the same indexation. The color coding is the same for both figures, and corresponds to the type of vertex. 

\begin{thebibliography}{15}
\bibitem{grunbaum} B. Gr\"unbaum and G. C. Shepard, Tilings and Patterns (Freeman, New York, 1987)
\bibitem{wangchen} N. Wang, H. Chen, and K. H. Kuo, Phys. Rev. Lett. {\bf 59}, 1010 (1987)
\bibitem{schneider2023} Emmanuel Gottlob and Ulrich Schneider, Phys. Rev. B {\bf 107}, 144202 (2023)
\bibitem{grimmreview} U. Grimm and M. Schreiber, in book Quasicrystals: Structure and Physical Properties, Ed. H.-R. Trebin, Wiley-VCH (2003)
\bibitem{jagaRMP} A. Jagannathan, Rev. Mod. Phy. {\bf{93}} 045001 (2021)
\bibitem{kalugin2014} Pavel Kalugin and Andr\'e Katz, J. Phys. A {\bf{47}} 315206 (2014)
\bibitem{mace2017} Nicolas Mac\'e et al, Phys. Rev. B {\bf 96} 045138 (2017) 
\bibitem{koga}  Akihisa Koga, Phys. Rev. B {\bf{104}} 115125 (2020)
\bibitem{mirza}  M. Mirzhalilov and M.O. Oktel, Phys. Rev. B 102, 064213 (2020) 
\bibitem{oktel} M. O. Oktel, Phys. Rev. B {\bf{104}} 014204 (2021)
\bibitem{grimmroemer} Uwe Grimm and Rudolf A. Roemer, Phys. Rev. B {\bf 104} L060201 (2021) 
\bibitem{tarzia} A. Jagannathan, P. Jeena and M. Tarzia, Phys. Rev. B {\bf 99} 054203 (2019)
\bibitem{bedaride} Nicolas Bedaride and Thomas Fernique, in Aperiodic Crystals, Eds. Schmid, S., Withers, R., Lifshitz, Springer, Dordrecht (2013); Discrete Comput. Geom. {\bf 54 } 980 (2015)
\bibitem{duneau89} M. Duneau, R. Mosseri and C. Oguey, J. Phys. A {\bf 22}, 4549 (1989).
\bibitem{duneau94} M. Duneau in Lectures on Quasicrystals Eds. F. Hippert and D. Gratias (Les Editions de la Physique, Les Ulis, 1994)
\bibitem{aj2023} A. Jagannathan, https://arxiv.org/abs/2304.04409
\bibitem{kelput} Johannes Kellendonk and Ian F. Putnam, in Directions in Mathematical Quasicrystals, Eds. M. Baake and R. Moody, CRM Monograph series Providence RI (2000)
\bibitem{bellissardgap} J. Bellissard, Lecture Notes in Phys., {\bf{153}}, Springer, (1982).
\bibitem{hauck} J. B. Hauck et al, Phys. Rev. Res. {\bf{3}}, 023180 (2021).
\bibitem{varjas} Daniel Varjas et al, Phys. Rev. Lett. {\bf{123}} 196401 (2019)
\bibitem{cao} Ye Cao et al, Phys. Rev. Lett. {\bf{125}} 017002 (2020)
\bibitem{ghadimi} Rasoul Ghadimi et al, Phys. Rev. B {\bf{104}} 144511 (2021)
\bibitem{takemori} Nayuta Takemori, Ryotaro Arita and Shiro Sakai, Phys.Rev. B {\bf{102}} 115108 (2020)
\bibitem{andrade} R. Araujo and E. C. Andrade, Phys. Rev. B {\bf 100}, 014510 (2019).
\bibitem{fukushima} Takumi Fukushima, Nayuta Takemori, Shiro Sakai, Masanori Ichioka, and Anuradha Jagannathan, Phys. Rev. Res. {\bf{5}} 043164 (2023)
\bibitem{huangLiu} H. Huang and F. Liu, Phys. Rev. B {\bf{98}} 125130 (2018)
\bibitem{deandrade}  E. de Andrade,  Anuradha Jagannathan, Eduardo Miranda, Matthias Vojta, and Vladimir Dobrosavljevic, Phys. Rev. Lett. {\bf 115}, 036403 (2015).
\bibitem{lin} C. Lin, P. J. Steinhardt and S. Torquato, J. Phys. Cond. Mat {\bf{29}} 204003 (2017)
\bibitem{sakai} Shiro Sakai, Ryotaro Arita, and Tomi Ohtsuki, Phys. Rev. B {\bf 105} 054202 (2022) 
\bibitem{trambly} G. Trambly de Laissardi\`ere, C. Oguey and D. Mayou, Phys.: Conf. Ser. {\bf 809} 012020 (2017)
\bibitem{wessel} S. Wessel et al, Phys. Rev. Lett. {\bf 90}, 177205 (2003)
\bibitem{szallas2007} A. Jagannathan et al, Phys. Rev. B {\bf 75} 212407 (2007)
\bibitem{ghosh} P. Ghosh, https://doi.org/10.48550/arXiv.2301.11331
\bibitem{flicker} S. Singh et al, https://doi.org/10.48550/arXiv.2302.01940
\bibitem{senechal} Marjorie Senechal, Quasicrystals and geometry, Cambridge University Press, 2009
\bibitem{deloudi} Walter Steurer and Sophia Deloudi, Crystallography of quasicrystals: concepts, methods and structures, Springer-Verlag Berlin and Heidelberg, 2009 
\bibitem{baakegrimm} Michael Baake and Uwe Grimm, Aperiodic order: a mathematical invitation, Cambridge University Presss, 2013
\bibitem{sirebell} Sire and J. Bellissard, Europhys. Lett. {\bf{11}} 439 (1990)
\bibitem{benza} V.G. Benza, C. Sire, Phys. Rev. B {\bf{44}} (1991) 10343.
\bibitem{zhongmoss} J.X. Zhong and R. Mosseri, J. Phys. I France {\bf{4}} 1513 (1994)
\bibitem{jagannathanHAFM} A. Jagannathan, Phys. Rev. Lett. {\bf 92}, 047202 (2004); Phys. Rev. B {\bf 71}, 115101 (2005).
\bibitem{henley} Christopher L. Henley in Quasicrystals: the state of the Art, Ed. D.P. di Vincenzo and P.J. Steinhardt, World Scientific, Singapore (1991)
\bibitem{jagapiech} A. Jagannathan and F. Piechon, Phil. Mag. {\bf 87} 2389 (2007)
\bibitem{socolar} J. Socolar, Phys. Rev. B {\bf 39} 10519 (1989)

\bibitem{grimm99} U. Grimm and M. Schreiber in Quasicrystals - An Introduction to Structure, Physical Properties and Applications, edited by J.-B. Suck, M. Schreiber and P. H\"aussler (Springer, Berlin, 2002)
\bibitem{zijlstra} E. S. Zijlstra, J. Non-Cryst. Sol. {\bf{334-335}} p. 126 (2004) 
\bibitem{godreche} C. Godreche, J.M. Luck, A. Janner and T. Janssen, J. Phys. I France {\bf 3}  1921 (1993)
\bibitem{siremossadoc} C. Sire, R. Mosseri and Jean-Francois Sadoc, J. Phys. France 50 (1989) 3463 (1989)
\bibitem{thiemschr} S. Thiem and M. Schreiber, Phys. Rev. B {\bf 85}, 224205 (2012)
\bibitem{jaga2013} A. Jagannathan and M. Duneau, Europhys. Lett. {\bf 104} 66003 (2013); N. Mace, A. Jagannathan and M. Duneau, Crystals {\bf 6}, 124 (2016)


\end{thebibliography}
\end{document}